Roadmap

# Roadmap on signal processing for next generation measurement systems

**Dimitris K Iakovidis**[1,*] , **Melanie Ooi**[2] , **Ye Chow Kuang**[2] , **Serge Demidenko**[2,24] , **Alexandr Shestakov**[3], **Vladimir Sinitsin**[3] , **Manus Henry**[3,5,6] , **Andrea Sciacchitano**[7] , **Stefano Discetti**[8] , **Silvano Donati**[9] , **Michele Norgia**[10] , **Andreas Menychtas**[11], **Ilias Maglogiannis**[11] , **Selina C Wriessnegger**[12] , **Luis Alberto Barradas Chacon**[12], **George Dimas**[1], **Dimitris Filos**[13,14], **Anthony H Aletras**[13,14] , **Johannes Töger**[13], **Feng Dong**[15] , **Shangjie Ren**[15] , **Andreas Uhl**[16], **Jacek Paziewski**[17] , **Jianghui Geng**[18], **Francesco Fioranelli**[7] , **Ram M Narayanan**[19] , **Carlos Fernandez**[20], **Christoph Stiller**[20], **Konstantina Malamousi**[1], **Spyros Kamnis**[21], **Konstantinos Delibasis**[1], **Dong Wang**[22] , **Jianjing Zhang**[23] and **Robert X Gao**[23]

[1] University of Thessaly, Lamia, Greece
[2] University of Waikato, Hamilton, New Zealand
[3] Sunway University, Bandar Sunway, Malaysia
[4] South Ural State University, Chelyabinsk, Russia
[5] Coventry University, Coventry, United Kingdom
[6] University of Oxford, Oxford, United Kingdom
[7] Delft University of Technology, Delft, The Netherlands
[8] Universidad Carlos III de Madrid, Leganés, Spain
[9] University of Pavia, Pavia, Italy
[10] Politecnico Milano, Milano, Italy
[11] University of Piraeus, Piraeus, Greece
[12] Institute of Neural Engineering, Graz University of Technology, Graz, Austria
[13] Lund University, Skane University Hospital, Lund, Sweden
[14] Aristotle University of Thessaloniki, Thessaloniki, Greece
[15] Tianjin University, Tianjin, People's Republic of China
[16] University of Salzburg, Salzburg, Austria
[17] University of Warmia and Mazury in Olsztyn, Olsztyn, Poland
[18] Wuhan University, Wuhan, People's Republic of China
[19] Pennsylvania State University, University Park, PA, United States of America
[20] Karlsruhe Institute of Technology (KIT), Karlsruhe, Germany
[21] Castolin Eutectic-Monitor Coatings Ltd, Newcastle, United Kingdom
[22] Shanghai Jiao Tong University, Shanghai, People's Republic of China
[23] Case Western Reserve University, Cleveland, OH, United States of America
[24] Massey University, Auckland, New Zealand

E-mail: diakovidis@uth.gr

CrossMark

## Abstract

Signal processing is a fundamental component of almost any sensor-enabled system, with a wide range of applications across different scientific disciplines. Time series data, images, and

* Author to whom any correspondence should be addressed.

video sequences comprise representative forms of signals that can be enhanced and analysed for information extraction and quantification. The recent advances in artificial intelligence and machine learning are shifting the research attention towards intelligent, data-driven, signal processing. This roadmap presents a critical overview of the state-of-the-art methods and applications aiming to highlight future challenges and research opportunities towards next generation measurement systems. It covers a broad spectrum of topics ranging from basic to industrial research, organized in concise thematic sections that reflect the trends and the impacts of current and future developments per research field. Furthermore, it offers guidance to researchers and funding agencies in identifying new prospects.

(Some figures may appear in colour only in the online journal)

## Contents

# 1. Introduction

*Dimitris K Iakovidis*

University of Thessaly

In the history of science, the establishment of signal processing as a discrete field of science is placed in 1940s [1]. In that decade, masterpiece papers were published, mainly in the context of communications, with a monumental work to be that of Claude Shannon's 'A mathematical theory of communications.' That period signified the progress of signal processing also in the context of other fields, including radio detection and ranging (radar) technology, which flourished upon the needs of military applications during World War II [2]. Since then, especially after the revolution of digital technology of the sixties, signal processing has become an integral part of almost any sensor-enabled system.

The applications of signal processing are numerous, extending well-beyond the domain of communications. Multisource temporal data series, two or higher dimensional data structures, such as images and video, can be regarded as signals. Signal quality is affected by different factors, such as the characteristics of the sensors used for their acquisition, and non-deterministic phenomena related to the data acquisition environment. Signal processing methods have been devised to transform the signals so that different application needs are met. Common transformations aim to signal quality or feature enhancement and signal compression. Signal processing methods may precede, or be an integral part, of signal analysis methods, aiming to reveal important information about the content of the signals, including their semantics and the measurement of observable quantities. It is therefore evident that there are strong dependencies between sensor-based measurements and signal processing, either in the preparation phase or in the analysis phase of the signals considered by a measurement system.

In the recent years, the increase of computational resources has triggered a remarkable progress on adaptive systems with generic architectures, enabling the solution of more and more difficult signal processing and analysis problems. Such systems are the basis of machine learning (ML) and artificial intelligence (AI), and today, complex architectures, such as deep artificial neural networks (DNNs) [3], may include millions of free parameters that can be tuned by an algorithm to infer solutions, based solely on the input data. This trend, usually referred to as *deep learning (DL)*, has already entered the domain of measurement science, with several works indicating the effectiveness of today's ML methods to solve measurement problems.

This roadmap presents a critical overview of the state-of-the-art signal processing and analysis methods and applications, aiming to highlight future challenges and perspectives towards next generation measurement systems. These are measurement systems of the Fourth Industrial Revolution, leveraging and contributing to the scientific and technological advancements of the next decades. It is organized in four sections. Section 2 identifies issues related to signal processing that are worth considering in contemporary measurement systems. These include uncertainty modelling, which is a fundamental issue that is still open, and issues related to networked multisensor measurement systems based on internet of things (IoT) technologies. Section 3 identifies challenges with respect to signal processing for optical measurement systems. In this direction signal processing perspectives are discussed in the context of particle image velocimetry (PIV) and interferometry. Sections 4 and 5 deal with a broad spectrum of applications, where measurement quality and efficiency can have a significant societal or economic impact. These span to biomedical, remote sensing, environmental, and industrial domains.

Many of the challenges and perspectives identified in this roadmap are associated with ML, which is undoubtedly a promising direction. However, it should not be considered as a panacea for the treatment of any signal processing or measurement problem. By taking a closer look to a problem under investigation and by understanding the involved physical processes other solutions, either without or combining DL with knowledge about the problem under investigation, could be more efficient. This is also highlighted in the roadmap, along with a multitude of other directions for novel research results and progress.

## 2. Signal processing considerations in contemporary measurement systems

### 2.1. Signal uncertainty modelling

*Melanie Ooi*[1], *Ye Chow Kuang*[1] and *Serge Demidenko*[2]

[1] University of Waikato
[2] Sunway University

### Status

Since the advent of the digital revolution, the volume of data acquisition and processing has been growing at an alarming rate. It has been fuelled by new technologies and tools as well as by the naïve confidence that there would be wisdom and knowledge within the acquired data that can reveal new insights. However, the confidence in data-driven decision-making does not depend on the amount but rather on the usefulness of data [4]. Data usefulness can be generally categorised into (a) discrimination of what is relevant, (b) interpretation of the information acquired, and (c) identification of sources, measurement, and management of uncertainties of the data (signals) as they inherently affect the operations of data-driven systems.

It is therefore crucially important to account for uncertainties when employing or designing such systems especially in mission-critical domains such as health, environment, security, etc. The sources of uncertainties are associated with the measurement of system input signals as well as with the processing of these signals by the system itself (that is naturally not perfect and having physical, technical, performance, and other constraints). The technique to infer output uncertainties given the uncertainties of input is known as the propagation of uncertainty. Known uncertainties of the inputs to a known measurement system can be propagated through a model of the system to find the probability density function or cumulative distribution function of the constraints, allowing characterisation of the uncertainties imposed by the system. This, in turn, facilitates various types of system analysis where needed. If no adequate model of the system is known, a model can be developed by measuring input quantities in relation to output quantities to determine their relationship.

### Current and future challenges

The *Guide to the Expression of Uncertainty in Measurement* (GUM) [5] outlines three methods to evaluate the propagation of uncertainty:

(a) GUM uncertainty propagation framework. This analytical approach is sufficient for largely linear cases. Unfortunately, it can be inaccurate in real-world scenarios in presence of non-linearities.
(b) Monte Carlo simulation. It is valid for wider classes of uncertainty estimation problems. However, it is computationally more expensive and requires large simulation sizes for complex problems thus limiting its use.
(c) Analytical methods based on statistical moments. In fact, the GUM uncertainty framework outlines the simplest form of the moment-based method with just two moments: mean and standard deviation. This somewhat limits the framework's application since just two moments are insufficient to model complex problems.

A concise yet quite comprehensive outline of the area of modelling uncertainty of signals can be found in [6] published almost a decade ago. There have been a large number of new results in this very active research domain including signal uncertainty evaluation, propagation through the measurement procedures, modelling, applications, etc, while further extending the foundations formulated by GUM, e.g. [7–10]. This area continues to expand while progressing to address the numerous challenges, such as advancing approaches for uncertainty evaluation for time-dependent measurements and their implementation for routine applications, increasing efficiency and reducing the cost of a Monte Carlo method for uncertainty evaluation, studying advanced approaches based on computing moments of higher orders of the output-of-interest, developing computationally efficient uncertainty evaluation techniques and tools enabling real-time applications, and so on.

### Advances in science and technology to meet challenges

Among the promising advances in the field under discussion is the analytical modelling of the uncertainty within a process of technical design optimisation. Uncertainty evaluation tools are employed to estimate the reliability and robustness of proposed design solutions. Here the reliability reflects the level of confidence in meeting a physical design constraint, whereas robustness refers to the sensitivity of an output to the uncertainties in the inputs. This method has been implemented in the structural design [11], whereby the overall reliability and robustness of a prescribed constraint of the structure comprising many elements (e.g. beams, columns, etc) is often sought-after based on the knowledge of uncertainty of each element. This information is then fed into deterministic optimization algorithms (i.e. where uncertainties of the design parameters are not considered) to find the best design solution while minimising weight/cost and meeting the physical constraints with a desired level of reliability along with less susceptibility to the system uncertainties. As a result, a design that meets the required reliability and safety can be achieved. This technique likely would be applicable to designing measurement systems with numerous sensor elements.

Modelling uncertainty propagation within a complex system will be an important tool to build a robust decision-making system. And herein lies the problem—the existing uncertainty propagation methods have historically been focused on techniques that serve the main role of reporting/estimating output uncertainty and are confined to static or quasi-static operating environments [12]. State-of-the-art techniques are capable of evaluating input-output uncertainty of systems represented by linear or polynomial functions. Theoretical advances to estimate input-output uncertainty of the systems representable by more flexible models such as radial basis functions or artificial neural networks (ANNs) would expand the ability to estimate uncertainty in complex decision making. In the cases where closed-form solutions cannot be found, limiting solutions (similar to the central limit theorem) could provide useful theoretical bound in the design of large and complex signal processing systems. Further investigations, development and introduction of such improved techniques will help to advance the research subject area of uncertainty propagation beyond the current boundaries.

Technical challenges for uncertainty evaluation and modelling are mainly associated with complex signal processing structures having a multitude of distributed inputs and outputs deployed to carry out long-term missions in varying environments (e.g. large-scale IoT systems) where the data reliability and system stability are not guaranteed. This would negatively affect the fundamental items of dealing with uncertainties— calibration (and system identification) and traceability [13]. A viable solution would be self-adaptive sensing with soft-calibration using the acquired data, where yet again, signal and uncertainty modelling would play a central role in developing efficient and robust self-tuning algorithms. Transferability between applications would be important to enable cost-effective large-scale integration. Improving data acquisition and sensor fusion have to be achieved as well as optimisation of the computational efficiency (e.g. achieving the best trade-off between computational accuracy versus resource utilization [14]) would also need to be advanced.

## Concluding remarks

It can be expected that the future intelligent measurement systems with embedded data-driven decision-making will continue to be characterised by the large-scale long-term deployment of multitudes of sensing elements connected to high-performance signal or data processing equipment. The uncertainty associated with the signals or data arriving from the sensing elements is to be accounted for along with the uncertainty of signal processing in time or complex digital computations performed by the system thus supporting the avoidance of unintended wrong decisions or results at the system outputs. And that is where lies the importance of the advancement of the theory, practice and tools for uncertainty modelling and evaluation (a good example of an uncertainty evaluation tool is given in [15]). The progress achieved in many topics associated with this field in recent years has been very significant. Yet, in terms of finding more general solutions that would be required for the design and deployment of the next-generation measurement systems, the challenges are serious.

## Acknowledgments

The authors would like to express appreciation for the support provided to their research by the IEEE Instrumentation and Measurement Society, Massey University, and Royal Society of New Zealand. The valuable contributions made by A Rajan, H Carstens, and R Zhang are acknowledged with thanks.

## 2.2. Signal processing for IoT-based measurements

*Alexandr Shestakov*[1], *Vladimir Sinitsin*[1] and *Manus Henry*[1,2,3]

[1] South Ural State University
[2] Coventry University
[3] University of Oxford

### Status

Recent decades have seen rapid developments in the scope and sophistication of networked sensor technology, particularly within the context of the internet of things (IoT) [16, 17]. The IoT raises a wide range of technical challenges for distributed sensing, including the development of energy-aware data acquisition systems, localization of mobile IoT nodes, synchronization protocols, and security [18]. Further questions arise concerning basic measurement functionality. The default assumption is that the measurement calculation and associated signal processing should be identical to that of an equivalent non-networked sensor. Perhaps, even simplified measurement calculations may be employed to reduce the complexity, cost and power consumption of the local device, if it can also be assumed that, at the network level, sophisticated data fusion may overcome the metrological limitations of individual sensing nodes. An alternative approach, promoted via an IEEE Recommended Practice [19, 20], would claim that 'a paradigm shift in the sensor world is on the horizon: the signal will be processed entirely at the point of measurement (POM)', i.e. that more, not less, measurement calculation should be performed locally, in order to reduce power, communication bandwidth and data storage requirements. The challenge is to develop schemes whereby local signal processing in individual nodes can be configured and modified, ideally as directed by the network level. For example, certain Fieldbus protocols, developed in the 1990s for industrial applications [16], provide a limited capability for downloading signal processing tasks into local nodes. A valve position can be modified using a control algorithm downloaded into a nearby temperature, pressure or flow sensor, for example. However, this capability is limited to distributing fixed function blocks (e.g. PID) across the device network. Here we present two promising techniques that provide more flexible, localised signal processing within the IoT domain. The first technique [20] provides a universal framework for extracting features from a measurement signal, resulting in a compact form of signal encoding. The second technique [21] provides a low cost, modular signal processing block that can be flexibly configured to perform a wide range of signal processing tasks, including for IoT devices.

### Current and future challenges

IoT roll-out will generate an exponential rise in the number of sensing nodes, deployed in diverse environments. This poses challenges to designers and users, whether of individual components or entire systems. 'As systems become more interconnected and diverse, architects are less able to anticipate and design interactions among components, leaving such issues to be dealt with at runtime. Soon systems will become too massive and complex for even the most skilled system integrators to install, configure, optimize, maintain, and merge' (cited in [21]). Big data and ML techniques have made remarkable progress in implementing top-down analysis, condition monitoring, and efficient operation for a wide range of complex systems. However, the question arises as to what signal processing capabilities might be provided at the lowest level to support autonomous and adaptive systems at higher levels, while also minimising processing and data bandwidth. Such developments would counteract the sometimes low priority currently afforded software development for IoT devices, where '… often software engineering is the last activity after mechanical and electrical design, facing a lack of information and limited development time because of delays in the other disciplines. On the other hand, bugs created in other disciplines need to be fixed by means of software' (cited in [21]). One route to new capability is to develop signal processing means to characterise real-time measurement behaviour through a compact form of coding, as a *further* stage of processing beyond the basic measurement. This should be simple enough to be implemented within any sensor node, while being sufficiently expressive to support a wide range of potential uses at both the local (sensor to sensor, sensor to actuator) and the system level. Another route is to consider whether sensor signal processing tasks can be characterised in a modular, parameterised form, so that task modification or augmentation can be compactly defined and communicated by the system down into the individual sensing node. This would facilitate flexible and adaptive processing of high bandwidth transducer data and produce high information content, low bandwidth, and possibly bespoke measurements, as required and requested by the higher-level system.

### Advances in science and technology to meet challenges

The IEEE Recommended Practice 21451 [20] provides an interpolation-based segmentation algorithm to encode any time series into a sequence of signal shapes. It is intended for real-time operation within any sensor and provides a standard means of signal identification and information fusion. Figure 1 shows the set of segment types, an example signal, and its corresponding coding vector: C is the character type, while M and T are timing markers. The technique excludes amplitude information, as it aims to encode only shape. However, maxima and minima are readily identified at the intersections of specific segment pairs—for example, maxima occur at 'de', 'df', 'ge', or 'gf' junctions. A simple application example [20] consists of heating liquid in a tank, where the heating actuator asks the temperature sensor to verify that the temperature is rising. The recommended signal shape characterisation is sufficient to support decentralised, ad-hoc interactions between devices, including the formulation of requests that can be interpreted and answered with low bandwidth yes/no responses. The Prism [21, 22] is a linear phase finite impulse response (FIR) filter where, unusually, the calculation is recursive so that the computational cost per sample is low and fixed, irrespective of filter length. Prism design and instantiation, given desired parameter values, is also trivial. This design simplicity and low computational cost support the use of Prism networks to carry out a variety of signal processing tasks, including low-pass, bandpass and notch filtering, and tracking [21], whereby frequency/phase/amplitude values are calculated for a sinusoidal, typically post-filtered, signal. Spectral analysis is also supported [23]. These properties enable the creation of new signal processing schemes within networked sensors on an ad-hoc, as needed basis. Figure 2 summarises a demonstrator: a wireless acceleration sensor [24] measures angular acceleration for condition monitoring of rotating machinery. A chipped or broken gear tooth results in modulation in the amplitude of the 4th harmonic of the rotation frequency; a Prism signal processing scheme can be instantiated for real-time tracking of this parameter to provide fault detection [25], where in this case the scheme is instantiated to support the current (assumed steady) speed of rotation.

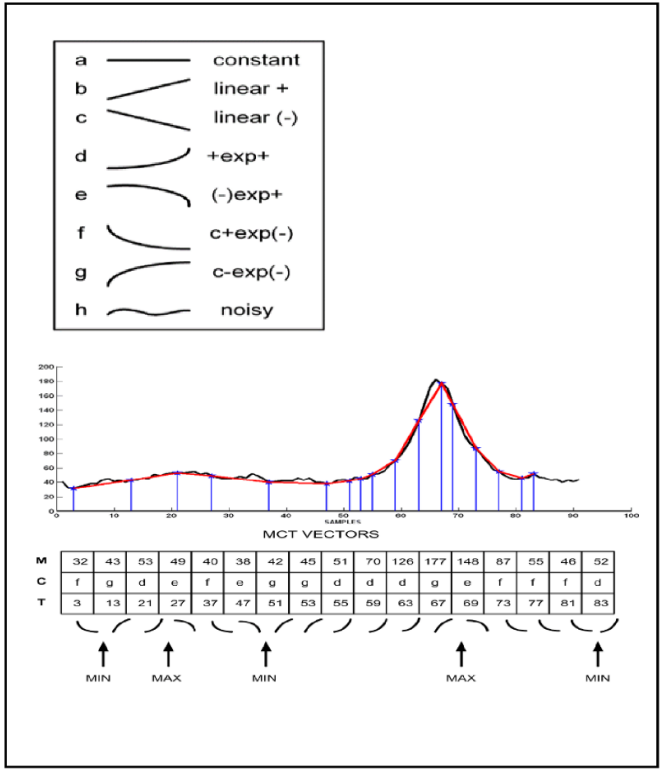

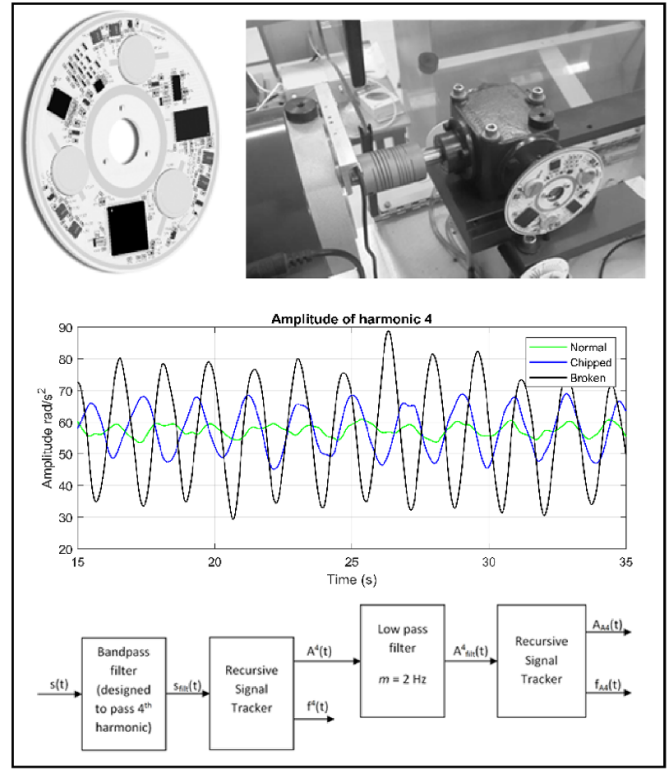

**Figure 1.** (Upper) classes for defining line segments; (lower) example of signal description using mark, class, tempos notation; © 2013 IEEE. Reprinted, with permission, from [19].

**Figure 2.** (Top left) wireless acceleration sensor; (top right) sensor mounted on gear box; (middle) time-varying amplitude of 4th harmonic for normal, chipped and broken gear; (bottom) Prism-based signal processing network to diagnose gear fault; © 2018 IEEE. Reprinted, with permission, from [25].

## Concluding remarks

A comprehensive approach to the evolution of the internet of things should include a reconsideration of how signal processing tasks can be defined and implemented. The development of novel techniques to facilitate flexible and adaptive calculations at the POM, i.e. within the sensing node itself, will help to reduce the computational, communication and data storage requirements that will continue to constrain the power and sophistication of fully autonomous systems. Current research typically considers, on the one hand, IoT implementations of well-established signal processing techniques, such as frequency estimation [26] or filtering [27], or on the other hand, application-specific requirements such as for healthcare [28, 29]. The two techniques described here suggest the potential for developing novel signal processing approaches to support localised, flexible signal processing as required by IoT. The examples provided are an encoding technique to characterise the shape of any signal, and the Prism, a recursive FIR signal processing building block with low design and computational cost. The diversity of these approaches suggest that alternative valuable techniques await discovery and application in future research programmes.

# 3. Optical measurements

## 3.1. PIV data processing

*Andrea Sciacchitano*[1] and *Stefano Discetti*[2]

[1] Delft University of Technology
[2] Universidad Carlos III de Madrid

## Status

PIV is a well-established versatile technique for velocity measurements of fluid flows. PIV is nowadays used in an impressively wide spectrum of applications, covering aerodynamics, biomedical flows, meteorology, oceanography, industrial applications and many more. The unique feature of PIV is its capability to deliver instantaneous velocity fields, thus enabling computation of vorticity fields and identification of coherent structures. Although with a wealth of variants, the processing of PIV images has reached a mature state. The core of the process is the use of algorithms to track the motion of particle images, in most cases based on either cross-correlation analysis [30] or individual tracking [31]. PIV algorithms have been progressively refined along the last three decades, leading to a sophisticated combination of image pre-conditioning and processing techniques [32] able to deliver a highly accurate flow characterization in an impressive range of conditions. Removal of unwanted background reflections from the image recordings is typically carried out by subtraction of the time-minimum or time-average pixel intensity. In most cases the velocity field is retrieved via cross-correlation analysis on small regions, often referred to as interrogation windows, covering the flow field (see figure 3). The process is generally multi-step, iterative, with progressive grid refinement. A detection, removal and correction of outliers is normally carried out after each pass. *A-posteriori* quantification of the measurement uncertainty is nowadays possible based e.g. on the analysis of the correlation statistics [33]. For three-dimensional (3D) flow measurements by tomography, the Shake-the-Box algorithm [34], based on the iterative particle reconstruction coupled with prediction and correction of the particle images positions along individual trajectories (see figure 4), has demonstrated higher accuracy and efficiency with respect to correlation-based algorithms.

PIV is now considered the dominant method in experimental flow characterization, and it is the most prominent candidate to provide full flow description in experiments. This is particularly relevant in applications where numerical simulations are either not reliable or not even feasible. While PIV processing can now be regarded as well established, large margins for improvement are foreseen with a clear reward ahead. PIV is still considered a technique for expert practitioners and with significant user-dependence on the results, being this especially true for volumetric measurements. Furthermore, we have witnessed the progressive increase in data size and dimensionality of PIV along its history, starting from few instantaneous snapshots of planar measurements to thousands of time-resolved volumetric representations of flow fields, each one including tens of thousands of vectors. The complete flow description is extremely attractive for validation of numerical methods and is fostering research efforts in data assimilation.

## Current and future challenges

Despite the widespread use of PIV in research laboratories worldwide, applications of the technique in the industrial environment are hindered by several limitations concerning both the measurement hardware and the data processing algorithms. Those include: the relatively small size of the measurement domain; the large times required for system calibration; the presence of unwanted laser light reflections, which locally hinder the evaluation of the flow velocity; the limited dynamic spatial and velocity ranges, both not exceeding O(100) [35, 36]; the dependence of the measurement results upon the user's expertise, and in particular the selection of the processing algorithm; the limited knowledge on the measurement uncertainty. These issues are further exacerbated when 3D flow measurements are performed, due to the use of multiple cameras and volumetric illumination.

Notwithstanding these difficulties, research efforts in the last decade have led to establish volumetric PIV as a standard tool for research [37]. Time-resolved velocity measurements are now available up to frequencies approaching 1 MHz [38], thus opening an unprecedented perspective for complete moderate-to-high-Reynolds number flow characterization. Nevertheless, the high frequency content of the velocity spectra is typically inaccessible due to measurement noise; currently, it is not clear whether multi-frame processing algorithms are capable to suppress the measurement noise without filtering the physical fluctuations. Enforcing first principles to extract field measurements of pressure [39] and scalar transport has pushed PIV in the last years beyond the mere measurement of kinematic quantities. Such approaches require high-repetition-rate equipment to achieve time resolution, thus limiting their applicability to relatively low speed flows. A current challenge is to overcome this limitation to transform PIV into a robust complete-flow-characterization tool in experimental aerodynamics.

## Advances in science and technology to meet challenges

Advances in high-repetition rate equipment, seeding particles for small response time, as well as multi-frame image interrogation algorithms, will be key enablers for MHz-range PIV [40], thus unveiling flow dynamics occurring at high-frequencies.

Bridging the gap between research laboratories and industry requires a higher level of automation in the PIV data acquisition and processing. Robotic PIV systems have been introduced [41] to enlarge the size of the measurement domain to cubic metres without compromising the data acquisition time. The development of AI approaches is considered crucial to optimise both image acquisition (automatic identification of viewing directions, measurement regions and number of samples to maximise the data quality) and data processing (user-independent optimisation of the processing parameters). Furthermore, robust uncertainty quantification methodologies based on the design of experiments framework shall be introduced that quantify the uncertainty associated with each of the relevant error sources [42]. Finally, image and data compression approaches will be crucial to enable researchers and engineers to handle the incredible wealth of flow information acquired.

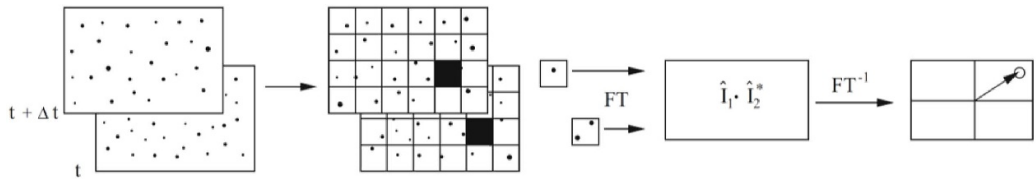

**Figure 3.** Analysis of double frame PIV recording based on digital cross-correlation. Reprinted by permission from Springer Nature Customer Service Centre GmbH: Springer. [35] © 2018.

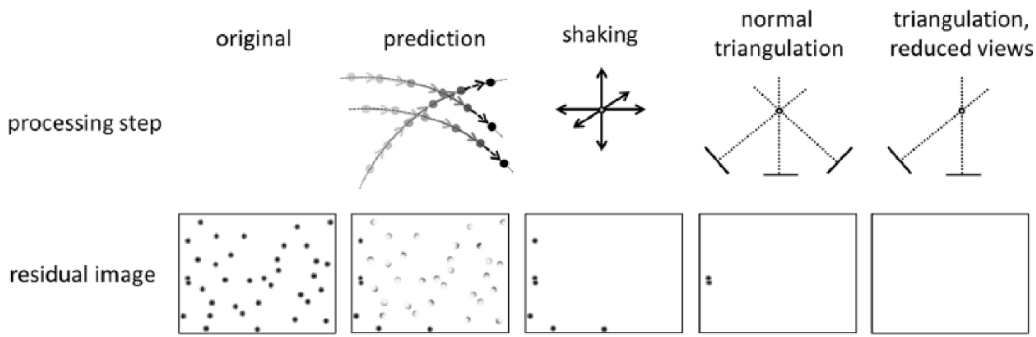

**Figure 4.** Schematic description of the Shake-the-Box algorithm. Reprinted by permission from Springer Nature Customer Service Centre GmbH: Springer. Experiments in Fluids. [34] © 2016.

Advances in data fusion and data assimilation approaches will have a key role in enriching the measurement data by enabling the determination of flow properties otherwise not accessible, such as surface pressure and wall shear stress, thus leading to the evaluation of the aerodynamic load distributions, or flow scales beyond the Nyquist limit imposed by the spatial sampling with particles. Promising pathways include imposing the compliance with the governing equations of fluid motion, either based on the Navier–Stokes or on the Lattice-Boltzmann formulations, and the use of variational data-assimilation frameworks employing adjoint computational fluid dynamics.

ML algorithms are also expected to be a key player in addressing the challenges to increase the robustness and user-independence of PIV image processing. Promising research avenues are opened by the deployment of DL algorithms. The perspective of such methods is to exploit the recent advances of computer vision to increase the temporal and spatial resolution of PIV beyond the limits of the image acquisition system. The developments in super-resolution ML algorithms (see, for instance, generative adversarial neural networks) are now offering intriguing research pathways. This is expected to stimulate research efforts on novel physics-informed algorithms since in most of the applications of PIV the underlying equations are known and can be enforced to regularize the final output. Furthermore, ML algorithms can pave the way towards embedding time-resolution in standard low-repetition-rate PIV using sparse high-repetition-rate simultaneous measurements, thus easing the achievement of a complete flow dynamics characterization.

## Concluding remarks

PIV is offering the perspective of a complete technique for flow characterization, with robust and reliable measurements to validate numerical simulations and to explore configuration where simulations are not accessible. The main challenges ahead are identified in the improvement of the flexibility and robustness of the technique to ease its transition from being a specialized technique to a standard tool also in industrial practice. Although hardware advances are foreseen as a strong beneficial asset for this process, very promising opportunities in this direction are offered by advanced signal processing. A strong embedding of first physical principles by means of data fusion, data assimilation or physics-informed DL methods is expected to be a key player in the next years. The recent advances in computer vision are also already contributing to setting up a new generation of PIV processing methods, opening new perspectives for improved accuracy, mitigation of the user dependence, and for uncertainty quantification.

## Acknowledgments

S D acknowledges funding from the European Research Council (ERC) under the European Union's Horizon 2020 research and innovation programme (Grant Agreement No. 949085).

## 3.2. Interferometry signal processing

*Silvano Donati*[1] and *Michele Norgia*[2]

[1] University of Pavia
[2] Politecnico Milano

### Status

Interferometry conveys information about distance and changes in distance by the interference of a sinusoidal light signal that is transmitted with its reflection from an object of interest, such as a mirror or a diffusing target. The interferometric signal to be processed comes from a photodetector; ideally the signal to be processed has the form:

$$I = I_0 \left[ 1 + m \cos \left( 2ks + \varphi_0 \right) \right]. \tag{1}$$

This is the output signal obtained at the photodetector as the beating of the reference beam $E_R = E_{00} \exp i\omega t$, and the measurement beam propagated forth and back to the target at distance $s$, $E_S = E_{S0} \exp i(\omega t - 2ks)$. We wish to trace back the distance $s$ or displacement $\Delta s$ contained in the phase term $\varphi = 2ks + \varphi_0$ under the cosine function. Specifically, $s = s(t)$ is the time-dependent distance of the target, $I$ is the photodetector current, $I_0 = E_{00}{}^2 + E_{S0}{}^2$ its mean value, $m = 2\,E_{S0}E_{00}/I_0$ the modulation index (or fringe visibility), $\varphi = 2ks$ the interferometric phase accumulated over the path $2s$, $k = 2\pi/\lambda$ the wavevector and $\lambda$ the wavelength, and $\varphi_0$ a constant phase term, equal to arm imbalance in a two-beam interferometer [43, 44]. While the problem is trivial when $s(t)$ is monotonic, because we can invert equation (1) as $s(t) = (1/2k)$ arccos $[(I/I_0 - 1)/m] - \varphi_0$, as soon as $s(t)$ has minima or maxima the cosine function periodicity (or the multi-values of arccos) prevent to obtain an unambiguous inversion.

Figure 5 shows an example of sine-wave displacement $s(t) = s_0 \sin\omega_0 t$ for which the signal $\cos(2ks_0 \sin\omega_0 t + \varphi_0)$ exhibits a slower periodicity in correspondence with the maxima/minima of $s(t)$ [43]. Traditionally, the problem is solved by modulating the laser wavelength (heterodyne interferometer, typical solution for vibrometers [45]), or by adding a second interferometric signal, of the type $\sin (2ks + \varphi_0)$, so as to have a complete pair $\{\sin\varphi, \cos\varphi\}$ identifying the phase $\varphi = 2ks$ unambiguously.

A different approach to overcome the ambiguity was developed after the introduction of the self-mix configuration of interferometry [46] by which, as soon as the feedback level is moderate the interferometric waveform becomes distorted and a switching appears with the sign of the increasing/decreasing displacement (figure 5, bottom panel). Taking advantage of this self-unwrapping, the inversion operation can be carried out with no error and a limit set by noise at about ±5 nm, provided the parameter of the experiment (feedback factor $C$ and linewidth enhancement $\alpha$) are known [47].

### Current and future challenges

The first challenge of interferometric measurements is that of applications to mechanical engineering [48] and machine tool control is to work on native, untreated surfaces (the non-invasiveness feature) and this means that the phase measurement shall be carried out in the *speckle pattern* regime [43] of optical field returning from the target. The speckle statistics corrupts both *amplitude* and *phase* of the returning field. In amplitude, we have fading of the signal, with the occurrence of dark speckles with near-to-zero amplitude, whereas the phase error amounts to a full $2\pi$ every time we go out of a coherence region or swing out of the speckle size. Amplitude fading is defeated by operating in space (or eventually wavelength) diversity, and the technique known as *bright speckle tracking* [49] has been shown to solve the problem, at the expense of a small addition to the optical objective collimating the beam out of the laser source: a piezo XY deflector moving the spot projected on the target so as to maximize the amplitude of the return. Once fading is overcome, it is the residual speckle phase error to limit the measurement accuracy. An analysis of the phenomenon [50] has revealed that, also in the speckle regime, we can attain a sub-μm resolution and precision by properly trading beam size and distance of operation. Even better is the hypothetic perspective of cancelling the speckle phase error by taking advantage of the Hilbert conjugation of phase and logarithm of the amplitude, till now demonstrated only partially and in special cases [43]. A second challenge to interferometry is the improvement of sensitivity and resolution limits for the most demanding applications, like the gravitational wave detection [51]. Figure 6 shows the optical scheme of VIRGO, the French-Italian detector that uses Fabry–Perot resonators (mirrors WE-WI and NE-NI) to improve sensitivity of a factor $\approx 10^{3,}$ equal to the finesse of the resonator, so as to reach the record sensitivity $\Delta s/s_0 = 10^{-23}$ that is necessary to sense gravitational waves.

A further challenge is the application to medicine and biology of low-coherence source imaging, also known as optical coherence tomography (OCT) [52]. Upon scanning the reference arm length, we get the in-depth scan of the surface from which the interferometric signal is collected, and in this way, we obtain a 3D image of the tissue (skin, blood vessel, retina) under test. The 3D image is further processed to identify and recognise the diagnostic details of interest (lesions, melanoma, etc). Since now, identification and recognition have been carried out by convolutional neural network (CNN) processing of the image at a single wavelength. A new possibility is offered by spectral-scanning OCT because it entails also the wavelength dependence of image details and therefore contains potentially more information as described in next section.

### Advances in science and technology to meet challenges

Signal processing for an interferometric instrument becomes increasingly crucial to tackle physical and technological limits. For mechanical applications, digital acquisition at high sampling frequency allows to better filter out disturbances and to highlight the desired measurement. Thanks to going digital, complex processing strategies, such as neural networks (NNs) and ML techniques, are applicable in real time to the phase evaluation (for example, NN are applied to gravitational waves). Various complex inverse problems can now be solved numerically, to reach the desired information.

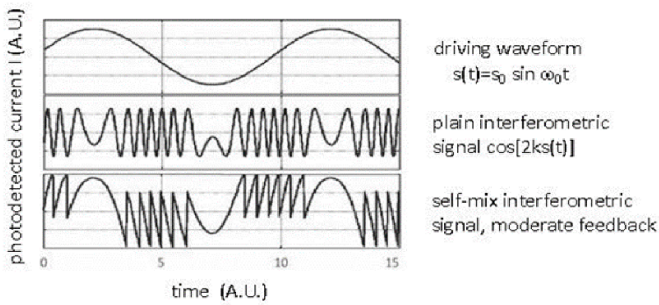

**Figure 5.** The interferometric signal has periodicities that make it difficult to trace back the actual displacement, like can be seen comparing the sine-wave excitation $s(t)$ (first panel), and the corresponding interferometric signal (second panel). Using a special configuration of interferometry, called self-mix, we get (in the moderate feedback regime, last panel) switching in the waveform telling the sign of the displacement $\Delta s$.

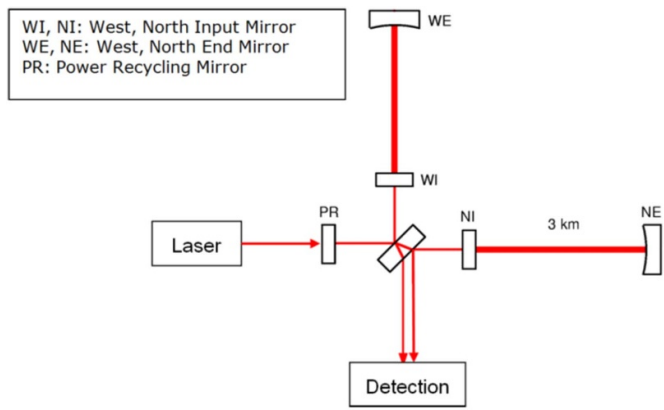

**Figure 6.** VIRGO optical scheme principle (from [51]). This is an example of very-large Michelson interferometer, with 3 km long arms and enhanced sensitivity. Reproduced from [51]. © IOP Publishing Ltd. All rights reserved.

As an example, the central problem of interferometric image reconstruction for detail recognition that are of clinical interest in using an OCT can be formulated in very general terms: it is to be able solving the constitutive Kubelka–Munk set of equation for the ongoing power flux $I(x, y, z, \lambda)$ and the backscattered power flux $J(x, y, z, \lambda)$ through the sample. Writing the fluxes simply as $I(z, \lambda)$ and $J(z, \lambda)$, i.e. letting the $x, y$ dependence implied, we have:

$$\mathrm{d}\,I(z, \lambda)/\mathrm{d}z = -A(z, \lambda)\,I(z, \lambda) + s(z, \lambda)\,J(z, \lambda)$$
$$\mathrm{d}\,J(z, \lambda)/\mathrm{d}z = A(z, \lambda)\,J(z, \lambda) - s(z, \lambda)\,I(z, \lambda) \qquad (2)$$

where $A(z, \lambda)$ and $s(z, \lambda)$ are the attenuation and backscattering coefficients of the details inside the image at depth $z$.

In the equation above, the quantity $J(z, \lambda)$ is the measurement output coming from the OCT interferometer, and our computational task is to find out the best estimates of the spectral attenuation $A(z, \lambda)$ and of the spectral backscattering $s(z, \lambda)$ coefficients for every pixel of the $x, y$ image, given the measured $J(z, \lambda)$. The problem is a difficult one because it is ill-conditioned and requires an algorithmic improvement of the calculation process respect to the well-known Gauss–Newton method, but is worth considering because the $A(z, \lambda)$ dependence more easily identifies the class of detail in the 3D image compared with the CNN processing used so far. The rather large dimensionality (the typical image may have 64 pixels in each coordinate and 16 values of $\lambda$) and the real time requirement add another challenge to the computation, which is however interesting because applicable to a wide class of problems.

## Concluding remarks

Interferometry is one of the most powerful measurement techniques in Physics, Engineering and Biomedical Sciences, and we have observed it is advantageously cross-fertilised by the advances in technology on one side and advances in data processing on the other side. Modern techniques of signal processing find broad application to interferometry, for phase evaluation or reconstruction of the desired measurand. The current signal processing trend shows a progressive evolution from analog systems [45] to digital processing (see bibliography of [44]), thanks to high-speed sampler with high-resolution, and to real-time signal processors. Next generation of interferometry could be simplified in optical architecture and more and more advanced in digital signal processing: this evolution will allow to reduce the cost, thus broadening the possibilities of application in the industrial world (see [43, 44, 53, 54]).

# 4. Biomedical measurements

## 4.1. Biosignal processing for pervasive health monitoring

*Andreas Menychtas and Ilias Maglogiannis*

University of Piraeus

### Status

Historically, healthcare is one of the domains that is highly benefited from the technological advancements in the different scientific areas, with the continuous creation of new clinical pathways, and innovative treatment plans. The contribution of information and communication technologies on this was fundamental, establishing the concepts of eHealth and mHealth, and accelerating the adoption of solutions for proactive and personalized care, which play key role in the enhancement of the mental and physical health and in the improvement of wellbeing in general [55]. In a broad context, 'eHealth' or 'electronic healthcare' refers to the sets of computing infrastructure and applications that assist the provision of medical services utilizing digitized medical data processed or in raw formats. In case the applications allow the users to be 'mobile' then we are referring to the 'mHealth' ecosystem. The evolution of sensing technologies, the availability of sophisticated hardware in commodity devices and the wide use of smart 'things', from phones, watches, and wearables, to advanced, special purpose equipment for the continuous measurement of vital signs, creates a rich profusion of data which can be exploited in pervasive health monitoring. The impact, in population and individual level, of these immense amounts of biosignal data which are produced every day, is even higher as a result of the modern methodologies and techniques which are used for signal processing and knowledge extraction. This knowledge, which is produced by analysing streaming data from biosignal sensors, and by correlating health data from personal, organizational, or public repositories, is nowadays crucial in the processes of decision-making by healthcare professionals and the assessment of patients' health condition. There are also cases, that the processing results are directly consumed by systems for the identification of health deterioration, estimation of risks, and for personalized coaching in the concepts of quantified-self and assisted living. ML, complex event processing and anomaly detection are among the technologies which are used for the processing of biosignals and are at the core of modern provisioning models in healthcare, such as the notion of digital-twin, the AI-assisted diagnostics, and the personalized medicine (figure 7). The importance of the vital sign processing approaches and the creation of valuable insights is also significant, if we take into consideration the disruption in everyday life due to extreme events such as the COVID-19 pandemic, and their effect in healthcare and other digital systems.

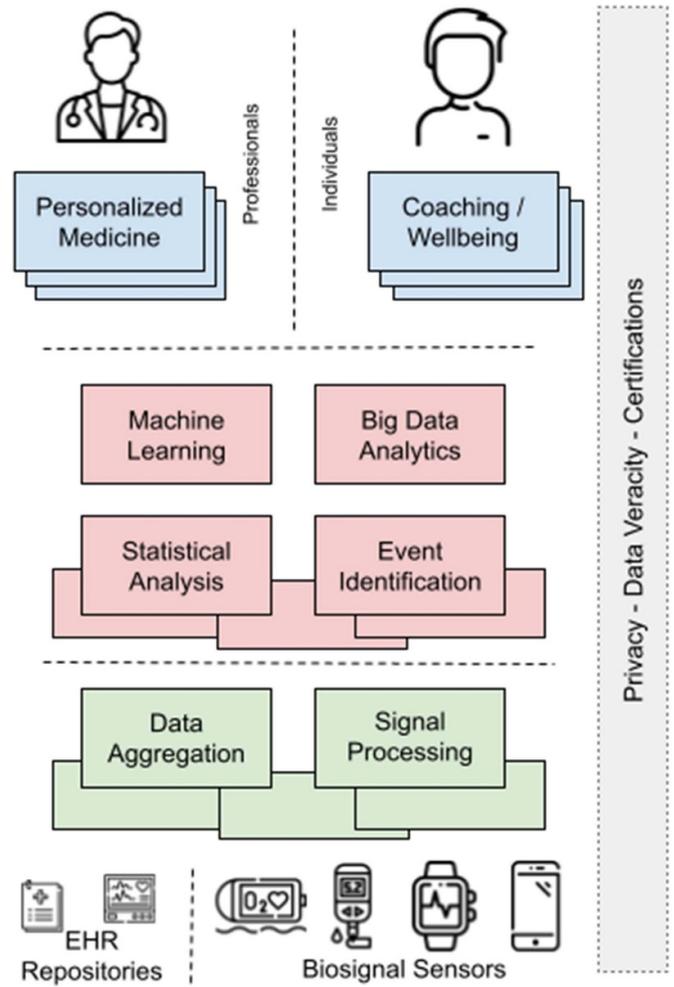

**Figure 7.** Vital sign processing overview for pervasive health.

### Current and future challenges

The main characteristic of the biosignals is that they are continuously updated and they can measure, quite effectively in most cases, important characteristics of patients' health status. At the same time, the nature of biosignal data itself introduces several challenges for their effective management, processing, and analysis. The constantly increasing number of different and diverse data types, following the abundance of hardware and sensors for measuring activity and biosignals, creates a complex environment for managing and manipulating the data [56]. The analysis processes and methodologies are becoming even more complicated taking into consideration the multi-modal nature of the signals, since they have different granularity, with sensors and data sources to produce on the one hand unprocessed raw data such as the users' heartrate, or aggregated data and events on the other hand, such as the identification atrial fibrillation following the measurement of ECG. The overall processing flow is hindered though by missing, untrusted and inconclusive measurements due to sensor misuse or network connectivity failures which introduce additional difficulties. da Costa *et al* [57] have provided a very enlightening analogy between the characteristics of health data from sensors and Big Data. The challenges related to the biosignal volume, velocity, variety, veracity, visualization and value are at the core of modern health applications which not only deal with data but also with the knowledge produced using processing techniques for noise removal, data fusion, semantic annotation and trend analysis. Since the processing typically takes place on mobile platforms or on the edge of the network, the similarities become apparent. The complexity increases in modern systems where different data streams, local and remote, medical and non-medical (e.g. location), should be combined and aggregated with historical data from the centralized electronic health records of healthcare organizations, taking also into consideration the limitations on the resources availability and energy consumption [58]. In this context, signal processing is often followed by analysis techniques which are taking place in multiple layers of the application stack, such as clustering, support vector machines, decision trees, neural and deep networks that are used for predictions, assessment of health parameters and risk estimation [59].

## Advances in science and technology to meet challenges

The first step towards the improvement of biosignal data processing and analysis is the unification of the techniques for manipulating data of different nature or data provided from different sources. The analysis methodologies may considerably benefit from the existence of an abstraction layer for data access which will be able to support all data types and sources across systems and devices. This should cover only the different characteristics of biosignals that can be found in electronic health records but also the interactions between systems that operate on the data, from mobile phones and edge devices to centralized digital repositories for vital signs and cloud computing platforms. Even though solutions for processing health data in different levels are discussed in literature [60, 61], their generalization and applicability for pervasive health monitoring is limited. Furthermore, processing methodologies dealing with streaming data such as complex event processing, or data from external, non-health related, sources such as users' location or other environmental parameters are even more difficult to be adopted without unified approaches for data exchange and health services interoperability. Although for several health data types there have been notable implementations of standards and models e.g. HL7 and FHIR [62], the biosignal data, and their extensions on Quantified-Self and social aspects [60], still lack of specifications that are natively accepted in processing and analytics frameworks. Example of this limitation is the incomplete list of Bluetooth GATT specifications which only cover a fraction of the constantly increasing list of biosignals acquired from devices and sensors. In addition, advancements on the semantic interoperability of biosignal data is a fundamental feature for their effective analysis from healthcare professionals and experts in different domains. This also includes the modelling and encoding of the analysis results and of the knowledge that is created in the separate processing levels so that experts and systems can effortlessly operate on them, and in that way transparently re-instantiate the data analysis loop following the annotation and fusion of new vital signs and health data on top of the knowledge that has been already created in the past.

## Concluding remarks

The creation of harmonized data lakes of biosignals on which the mechanisms for processing and analysis of data operate, should be the first step towards the wide adoption of advanced and efficient biosignal analysis methodologies for pervasive health monitoring. The increasing usage of sensors, the innovations in sensing technologies, and the advancements in fields of signal processing, event identification and knowledge extraction, require the use of models and standards which will facilitate the data integration processes. Correlation of biosignals and other health data, is fundamental in a ubiquitous healthcare ecosystem, where advanced analysis mechanisms can effectively function, allowing professionals to exploit them to gain insights on public health, and individuals to improve their wellbeing [63]. The landscape is changing fast, with experts and end-users embracing these technologies and concepts however, important steps are still required to ensure the veracity of vital signs, as well as the privacy of the users, through strict certifications and policies.

## 4.2. Electroencephalography (EEG) sensors and signal processing

*Selina C Wriessnegger and Luis Alberto Barradas Chacon*

Institute of Neural Engineering, Graz University of Technology

### Status

After the first recordings of electrical activity in humans by means of electrodes attached to the scalp by the German psychiatrist Hans Berger in 1924, a lot of research was carried out improving the signal acquisition [64]. Although the principles and basic procedures of non-invasive EEG have hardly changed, new advances in materials and electronic systems technologies support the development of a new generation of EEG sensors (figure 8). The most conventional sensors are wet electrodes utilizing a saline or different type of gels to increase signal to noise ratio (SNR) by increasing the scull contact area and decreasing impedance. Although they guarantee a high-quality signal recording they have several disadvantages. Generally the montage time is relatively long, skin irritations can occur and the hair must be washed after each measurement. To overcome these problems a new generation of dry electrode has been developed and evaluated by different companies (figure 8) and researchers [65–67]. For example Hinrichs *et al* [68] showed in a very recent study that the signal quality, ease of montage set-up and high usability of the dry electrodes comply with the needs of clinical applications. Even though the signals recorded from dry electrodes are sometimes noisier [69–71], their advantages of a fast setup, user-friendliness and wearer comfort are indisputable.

Several studies confirmed that the signal quality of dry electrodes can match the quality of wet electrodes depending on the context. For basic research of brain activity large multi-channel settings are required to apply sophisticated processing methods and deliver reliable insights in brain functioning. But nevertheless dry electrodes are successfully used in clinical studies [69] or in BCI applications [64, 69] with a limited number of channels. Both sensors are commercially available as wired or wireless systems. Especially the latter can be used out of the lab measuring free movements of persons without cable restrictions. Although wet electrodes are still considered as the golden standard, new advances in dry EEG electrodes give rise to extended future applications in more diverse research fields [70].

### Current and future challenges

When working with EEG signal processing, four major steps can describe the general pipeline: acquisition, pre-processing, feature selection, and modelling (figure 9). Each step has its own challenges: while recording, a small SNR and the reduction of different artefacts caused by environmental or physiological sources are amongst the greatest challenges in EEG studies. A second factor restricting common use of non-invasive EEG devices is the practical use of wet electrodes.

Although dry electrodes provide a faster setup and user comfort, they are still very sensitive to noise. For this reason, effort is also being made into developing active electrodes that preamplify EEG signals [72].

Pre-processing is an essential step in EEG data analysis as it is carried out to remove any artefacts and leaving only the desired EEG features for further analysis. The main challenge at this step is the reduction of noise while simultaneously keeping the relevant features, and removing those that will not be included in the model. This process can be as simple as channel selection, re-referencing methods, or frequency filtering, or more complex with blind source separation, wavelet transform (WT) methods, empirical mode decomposition, canonical correlation analysis or nonlinear mode decomposition [71]. The main challenge when extracting and selecting features from EEG signals is the incredible amount of variables that can be inferred from them. Features can be as simple as amplitude, statistical measures or segments used for event related potentials, but can also be abstract mathematical representations, like Hjorth features, differential entropy, higher order crossing, independent and principal component analyses, autoregressive, wavelet packet decomposition, or connectivity indices. This process is so overwhelming that some researchers use stochastic methods like genetic algorithms to sample a small, but relevant number of features to model. Modelling refers to the understanding of the neurophysiological bases for the behaviour measured. The mathematical and computational methods to understand how those bases relate to the physical measurement of EEG restrict our capacity to abstract and predict accordingly. Generally, simple models tend to be linear, but more complex models can make use of nonlinear Bayesian statistics, clustering algorithms, like nearest neighbours classifiers, NNs, or a combination of these, into ensemble algorithms. Many methods have been created over the years for specific use-cases, but no general-use model exists for EEG processing [73].

### Advances in science and technology to meet challenges

Several studies in the past compared dry-electrode performance with different types of wet electrodes (active and passive, water-based) but lacking homogeneity in comparison. This is one important step which should be addressed in future studies being able to make reliable statements about the signal quality of dry electrode approaches and pave the way for improvements. The creation and use of diverse electrodes allows for the development of new techniques for EEG measurement. This is the case of textile and tattoo electrodes for wearable technology, which allow for more diverse kinds of measurements. Conductive dry electrodes can be custom designed and 3D printed. Textile electrodes allow for soft wearable devices for prolonged measurements where participants can move freely. Tattoo electrodes provide minimally invasive and comfortable long-term measurements. Non-contact electrodes provide the promise of non-invasiveness, a fast setup, user-friendliness, comfort, and daily use electrodes. But also the combination of two different types of sensors, e.g. dry and wet, opens new ways to improve the SNR during EEG recordings [65, 72–74]. Although it is well known that the signal quality is strongly associated with the compatibility between the amplifier and electrodes, most of the existing literature has focused on electrodes only instead of the entire system. Therefore in future studies electrodes and assigned amplifiers should be evaluated together to find bottlenecks.

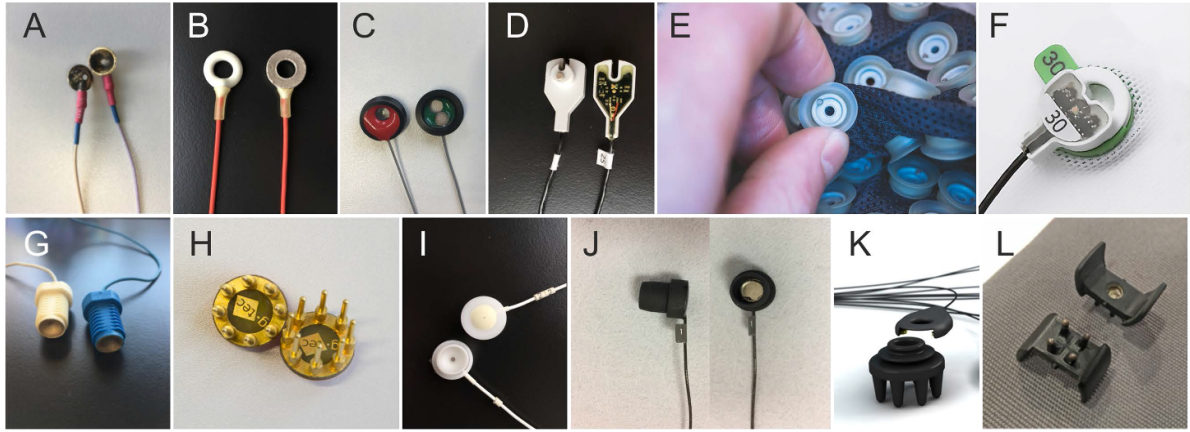

**Figure 8.** Different types of electrodes (1) passive sintered AgCl electrode, (2) gel-based active Ag/AgCl electrode (g.LADYbird from g.tec), (3) gel-based active electrodes (Brain Products GmbH), (4) gel-based wavegard electrodes (ANT Neuro), (5) gel-based slim active (actiCAP, Brain Products), (6) passive dry electrode (g.SAHARA electrode, g.tec), (7) water-based passive electrode (Mobita, TMSi), (8) dry EEG comb electrodes (OpenBCI), (9) unicorn hybrid electrode (g.tec), (10) semi-dry saline based electrodes (Greentek). Sensors 2, 6, 7 have been evaluated in [68]. Reprinted by permission from Springer Nature Customer Service Centre GmbH: Springer. [64] © 2021.

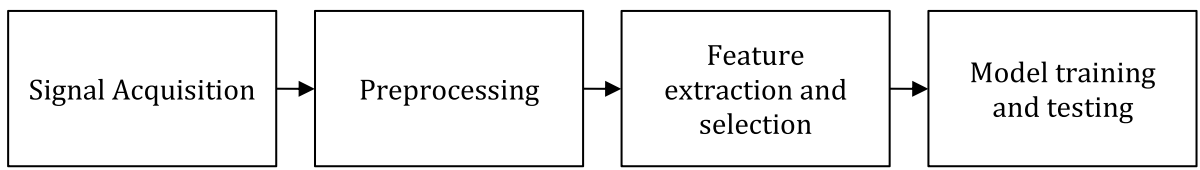

**Figure 9.** A generalized EEG signal processing pipeline.

DL provides techniques for data-driven discovery of relevant signal features, while being specifically resilient to noise. DL methods are capable of automatically learning features relevant to researched conditions, creating at the same time a model of the relationships between those features. For this reason, they can achieve similar results in problems such as mental state detection, motor decoding, or speech decoding [75]. A disadvantage of this automatic learning is the explainability of the created models, an area of opportunity in the field that has begun to be explored just recently [76]. This enables a more direct comparison of DL model with those created by traditional methods. A less explored advantage of DL model relies on transfer learning (TL); the possibility of using models that have been trained on specific datasets for different goals, or in different datasets, and being able to obtain accurate results with very little or no training. This is a good indicator that multi-objective multi-modal models are viable. The fast advances in computing hardware and software for DL make it one of the most promising techniques for reliable automated EEG data processing [77].

**Concluding remarks**

Summarizing, there are already a lot of different EEG sensors commercially available but not all of them deliver reliable and high signal quality recordings. Therefore a lot of research and development is still necessary to improve SNR, design features and usability of the sensors and amplifiers. The methods that drive this technology are bound to general signal processing techniques, but the continuously increasing computer power, AI, and the techniques that develop from it can be easily applied both as a modelling, and as a knowledge discovery tool. In the next few years, we expect a wide variety of user-friendly devices, combined with powerful computing techniques to improve the development and understanding of human EEG.

### 4.3. Optical image processing for *in vivo* measurements and diagnostics

*Dimitris K Iakovidis and George Dimas*

University of Thessaly

#### Status

Minimally invasive examination and treatment are continual challenges for medical science. In this context, optical imaging technologies play an important role by offering tools for *in vivo* screening at different wavelengths. Measurements performed *in vivo*, include the measurement of size, distance, and optical biomarkers. Endoscopy is the basic tool that can be complemented by emerging techniques revealing tissue properties valuable for diagnostic purposes, such as OCT, near-infrared fluorescence imaging, and multispectral/hyperspectral imaging [78]. Size measurements in endoscopic procedures are still based mainly on empirical comparisons of the target with a reference object, e.g. biopsy forceps. This approach can affect clinical decisions and lead to inappropriate recommendations [79]. Towards more objective and accurate *in vivo* size measurements, methods based on image processing have been proposed to perform comparisons with reference objects or structured light projections [80, 81]. To alleviate the need for external references, more recent image processing methods are based on ANNs and geometric calculations upon the optical camera characteristics, using one or more images as input (figure 10) [82, 83].

*In vivo* distance measurements are useful for the localization of findings within the body, e.g. anomalies. They are usually performed with wearable sensor arrays or radiologic procedures; however, recent image processing methods promise accurate results, more conveniently and less costly [83]. These methods follow the principles of visual odometry, and they have been investigated using both geometric and ANN-based approaches. Both distance and size measurements can be affected by the depth estimation accuracy. Typically, depth is assessed using two cameras in a stereoscopic setup, rendering the imaging system more costly and imposing constraints in miniaturization. Recent studies indicate that monocular depth measurements can be competently performed *in vivo* by DL methods [82, 84]. DNNs have dominated computer-aided diagnosis by offering a generic approach to image feature extraction. They have also proved effective in the detection of pathologies based on optical biomarkers [85]. New avenues to the identification of such biomarkers open with hyper-spectral/multi-spectral image analysis, which enables rapid clinical assessment of pathologies (figure 10) [86].

#### Current and future challenges

A major challenge for the development of trustworthy *in vivo* measurement systems based on image processing is the validation of the performed measurements. Since such measurements are performed within living organisms, it is difficult to establish a gold standard for accuracy and uncertainty estimation. This challenge becomes even greater considering the elasticity, the deformability, and the motility of the living tissues.

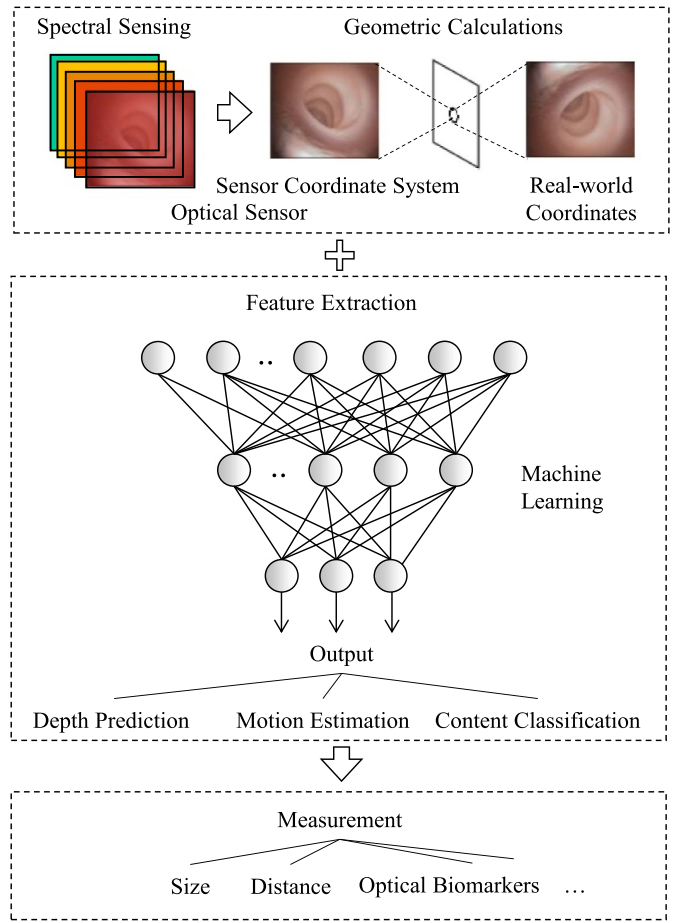

**Figure 10.** Architectural outline of a contemporary image-based measurement system.

Methods based on 'black box' approaches that learn from data, such as DNNs, have dominated image processing. DNNs are recognized for their high learning capacity; however, they usually require large volumes of training data to provide sufficient generalization. One of the largest obstacles today for the translation of the respective research to clinical practice, is the limited data availability, which originates: (a) from the vague and inflexible ethico-legal data sharing frameworks; and (b) from the need for image annotations by medical experts, which is usually a time-consuming and costly process. Another obstacle is that the 'black box' approaches are not explainable, in the sense that the clinicians cannot understand the reasons for the inferences of a ML system. Efforts dealing with this challenging issue are still limited [87].

The accuracy of image-based *in vivo* measurements depends on image quality. Restrictions with respect to image resolution are usually imposed by the small size of the image sensors satisfying the miniaturization requirements of the endoscopes, and the small energy consumption of battery-based solutions, e.g. wireless capsule endoscopes. Image quality can be significantly affected by noise artefacts, which are often induced by the image communication channels and lossy image compression algorithms, used to reduce the bandwidth and storage requirements. Body fluids or other content, e.g. mucus or debris, that may randomly interfere with the content of interest in the camera field of view can also be considered as noise affecting the quality of the measurements. However, coping with such noise, is a challenge that usually requires a different image processing approach in different medical imaging modalities.

Another challenge that is rarely addressed is the capacity of image-based measurement methods to provide coherent and consistent results on different images, acquired from the same or different imaging systems, e.g. to obtain comparable measurements from consecutive endoscopic video frames, or from the measurement of the same target using different endoscopes.

The majority of the current endoscopic image-based measurement methods have been proposed mainly in the context of gastrointestinal endoscopy [85], whereas fewer have been proposed in other contexts, such as laparoscopy [88] and colposcopy [89]. Challenging perspectives arise with respect to the application, adaptation, and validation of the current methods to other optical endoscopic imaging contexts and modalities.

## Advances in science and technology to meet challenges

A common approach to the validation of *in vivo* measurements is *ex vivo* or *in vitro* experimentation; however, usually it is costly, and it requires mechanical engineering skills to develop the experimental setup. *In silico* modelling and simulation can provide a sufficiently realistic solution for the implementation of digital twins of the measurement systems. Such a solution can combine different models, e.g. multiphysics, appearance, and device models, it can be applied for automated simulations on virtual patients, considering the material properties and the motility of the living tissues. Furthermore, it can complement the development *ex vivo* experimental setups via 3D printing. In the same spirit, realistic synthetic medical images with ground truth annotations can be generated to augment the training of DNN-based image processing methods (figure 11) [90–92]. However, the use of synthetic training data is only a compromise, which could be fundamentally solved by the adoption of a standardized digital image annotation and reporting framework by hospitals, and a research-friendly privacy preserving medical data sharing framework.

From a methodological perspective, the requirements of DNN-based image processing methods for annotated data, can be limited by investing more research effort on less supervised and unsupervised learning methods. For example, semi-supervised methods can be trained with both annotated and non-annotated data, weakly supervised methods can be trained with less informative annotations, and unsupervised methods can be supported by domain knowledge instead of training data. In this light, the detection and quantification of optical biomarkers could also be based on outlier detection methods, e.g. one-class classification systems, since the training of such methods can be based on normal samples.

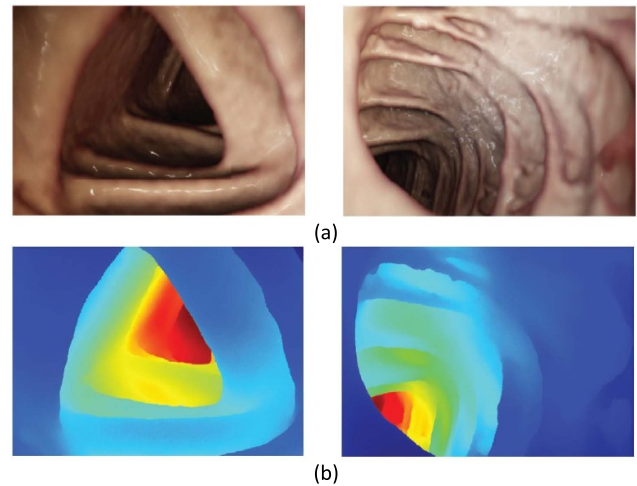

(a)

(b)

**Figure 11.** (a) Synthetic endoscopic images of the colon created from computed tomography (CT) images using cinematic rendering. (b) Respective ground truth depth maps. Reproduced from [90]. © IOP Publishing Ltd. All rights reserved.

Research directions towards image quality enhancement, include the development of improved super-resolution methods that may be able to offer even sub-pixel accuracy in measurement applications, less lossy or lossless compression algorithms, and noise reduction methods. The development of more robust image calibration, image matching and registration methods, can contribute to the improvement of the consistency and coherence of the image-based measurement results. To this end, illumination invariance and patient-specific or intra-patient image normalization methods are directions that require further investigation. An uncertainty analysis involving different patients and optical imaging devices, where possible, would contribute to a more essential progress in this field.

## Concluding remarks

Current advances on image processing and analysis indicate the feasibility of accurate image-based contactless *in vivo* measurements for minimally or even non-invasive optical imaging systems. Several studies have demonstrated their advantages over other methods based on special sensors or external references; however, most these methods are still at a relatively early research stage. Their translation to clinical practice requires efforts mainly towards enhancing their robustness, in a data-efficient way, and testing their performance in environments that enable a closer approximation of the real-world *in vivo* conditions. Addressing the challenges identified in this paper, will contribute to the evolution of next generation *in vivo* optical measurement systems that will be safer and less discomforting for the patients, as well as more reliable and efficient for the medical professionals.

### 4.4. MRI: signal processing and simulation

*Dimitris Filos*[1,2], *Anthony H Aletras*[1,2] and *Johannes Töger*[1]

[1] Lund University, Skane University Hospital, Department of Clinical Sciences Lund, Clinical Physiology
[2] Laboratory of Computing, Medical Informatics and Biomedical–Imaging Technologies, School of Medicine, Aristotle University of Thessaloniki

### Status

MRI is used in medicine for imaging, among others, anatomy, function, perfusion, diffusion and tissue viability which serve as a diagnostic tool, for risk assessment and for guiding treatment. MRI is based on the response of the hydrogen nucleus to radio frequency (RF) pulses while the patient lies inside a strong static magnetic field and its gradients. Different biophysical properties lead to image contrast used for diagnosis. To obtain conventional quantitative tissue parameters, multiple images are acquired with different scan parameters that affect the quantity to be measured. The images are then interrogated on a pixel-by-pixel basis to estimate the tissue parameters, such as T1, T2, T2*, diffusion coefficients, and extracellular volume. Accuracy and precision issues in tissue parameter mapping across MRI scanners have been documented. For example: pulse sequence, scan protocol, hardware characteristics, reconstruction software and parameter estimation methods may affect T1 and T2 mapping in cardiac MRI [93].

Ideally, it would be desirable to use a simple pulse sequence with a simple image reconstruction and parameter estimation method that can quantify a large number of tissue parameters on a variety of hardware platforms. In conventional MRI this is not possible since signal localization and tissue parameter estimation are separated into two distinct steps and treated as two different problems. Unifying the spatial localization and parameter estimation problems would potentially allow for more flexibility in executing faster scans and estimating more than one tissue parameters with less-than-ideal hardware and a simpler pulse sequence.

MR physics simulators are essential for solving the large-scale computational problem of a unified approach to signal localization and parameter estimation. To date, MR simulators are mostly used as research tools for pulse sequence design, hardware *in-silico* testing and educational purposes. Adding complexity to MR physics simulators to include hardware imperfections will allow for their use as part of both the image reconstruction chain and the parameter estimation process.

Several MR simulation platforms have been described in the literature. However, the increased complexity of realistic MR experiments leads to unacceptable simulation times thus hindering their applicability to clinical practice. Simplifications of the MRI physics models may have the potential to reduce the simulation times, but these come at the cost of reduced accuracy and generalizability. On the other hand, cloud computing and advancements in graphical processing unit (GPU) technology have led to the implementation of advanced MR simulation platforms, which are able to execute complex MRI simulations within reasonable times, which may enable clinical applications [94].

### Current and future challenges

AI and machine learning (ML) provide unprecedented opportunities for improving image-based diagnosis, prognosis, and risk stratification. However, these algorithms require a large amount of training data and preferably annotated datasets, which may be expensive to produce. In order to overcome this limitation, MRI simulations have been used to produce training datasets using realistic models of human anatomy [95]. This approach, which is an example of solving the *forward problem*, is based on the solution of the Bloch equations which yield the MRI signal that describes the evolution of the magnetization with a specific tissue model. The simulation protocol can incorporate various scanning parameters and recreate artefacts, if needed. The production and incorporation of more detailed anatomical models, such as statistical atlases that represent normal and pathological conditions, constitutes one of the main challenges.

A current and future challenge in MRI is solving the *inverse problem*, i.e. to estimate tissue parameters throughout the imaging volume based only on the acquired signal. MRI physics simulators are essential for solving the inverse problem since they provide a reference for comparing the acquired signal in an effort to obtain accurate reconstructions of quantitative multi-parametric maps. To this end, magnetic resonance fingerprinting (MRF) uses a specialized pulse sequence with pseudorandom pulse sequence parameters, such as RF flip angle and repetition times, in order to produce 'fingerprints' for each combination of tissue properties [96]. A sophisticated MR simulator has been used with the SQUAREMR framework to quantify T1 in cardiac MR using existing clinical pulse sequences [97]. More recently, Magnetic Resonance Spin TomogrAphy in Time-domain (MR-STAT) [98] was proposed, which considers the problem of parameter map reconstruction as a large-scale nonlinear inversion problem based on an MRI simulator, which can be solved using optimization algorithms (figure 12). This enables accurate tissue parameter estimation but requires long simulation times, on the order of hours for a single 2D brain slice even when using high-performance computing (HPC) clusters.

Simulation-based reconstruction algorithms have also the potential to measure additional tissue properties. Recently, an approach called MRF with exchange was proposed [99] to quantify sub-voxel relaxation times in single scan and extracellular volume, used as a marker of myocardial fibrosis. DL and MR simulators can also be used in tandem to better estimate tissue relaxation times [100].

The simulation-based reconstruction techniques have shown promising results, however, the transition from research to clinical practice remains challenging. Towards this goal

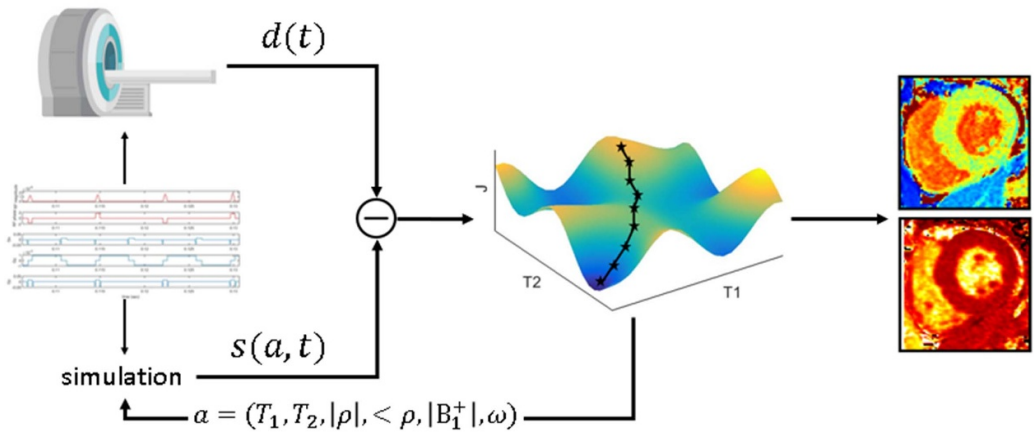

**Figure 12.** Graphical overview of the MR-STAT reconstruction algorithm. The signal $d(t)$ acquired by the scanner is compared with a simulated signal produced using the same pulse sequence in an anatomical model with known tissue properties $s(a, t)$. Optimization algorithms are used to change these properties so as to minimize their difference (error) and thus to reconstruct the images with the correct tissue characteristics.

both small-scale technical and large-scale clinical validations must be considered.

## Advances in science and technology to meet challenges

The main barrier for applying simulation-based techniques in clinical practice is the long computational times. Simulation-based reconstruction techniques based on inverse problem solution, such as MRF and MR-STAT, require fast and accurate computation of the simulated MR signal. Towards this goal, technological advances should aim to reduce simulation times.

Hardware innovations, such as HPC and general-purpose computing on GPU, enable the faster and parallel computation of the constantly growing complex models, allowing for performance improvements of computationally intensive calculations regarding tissue characterization. The integration of newer technologies has shown to progressively decrease the computational time.

Apart from raw computational power, which will always be useful, the implementation of novel algorithms can also help propel these methods into clinical practice. DL techniques have demonstrated success in solving medical imaging problems. As described by Liu *et al*, recurrent NNs can be used as an alternative to the Bloch physics simulation for computing large-scale MR signals and derivatives [101] and accelerate computation.

However, an ever-increasing amount of data requires increased computational power. In the future, quantum computers will likely outperform classical computer architectures and out stage them for applications requiring HPC. However, the design of quantum circuits differs significantly from that of a classical computer and thus the implementation of the NNs may require more qubits or lead to inconsistencies. Jiang *et al* proposed the QuantumFlow framework which is based on the co-design of NNs and quantum circuits [102] and could be used for processing high resolution MR images.

Smarter algorithms are needed for estimating the derivatives of functions that are used for optimization purposes. This will help reduce the overall computational load. Smarter algorithms are also needed for simulating MRI physics. Such an example has been demonstrated by prolonging the smallest simulation time interval [103]. In the end, smarter algorithms may allow for large speedup simulation time.

## Concluding remarks

Current MRI scanners are characterized by their increased cost and the fact that they are difficult to use, since a specialized technician with medical physics background is required to apply any modification related to the scanning protocol. Next generation systems must overcome these obstacles. Simulation-based image reconstruction approaches in MRI have demonstrated their early usefulness on tissue parameter estimation with high accuracy. The common ground challenge in this field is the need for reducing simulation times since the simulator is part of the inner loop of the algorithms. In addition, this approach can allow the reconstruction of numerous images simultaneously, such as diffusion, perfusion, or flow images, without the need of specialized pulse sequence to be applied. Advances in technological infrastructure and computer science and in particular GPU technology may lead to simulated experiments reflecting the full complexity of MR physics to increase accuracy of quantitative parameters. Feasibility studies and extensive clinical tests must be performed in order to ensure the usefulness of these methods. The standardization of the simulation-based reconstruction algorithms should be considered in terms of their application in clinical practice [104]. Finally, hardware advances can lead to reduction of the costs and thus the more extensive adoption of MRI in the clinical pathway.

## Acknowledgments

This research received funding from the Hellenic Foundation for Research & Innovation (HFRI), the Faculty of Medicine at Lund University, Sweden, the Region of Scania, Sweden, the Swedish strategic e-science research program eSSENCE, and Swedish Research Council Grant 2018-03721.

## 4.5. Ultrasound imaging: beyond conventional imaging and processing techniques

*Feng Dong and Shangjie Ren*

Tianjin Key Laboratory of Process Measurement and Control, School of Electrical and Information Engineering, Tianjin University

### Status

Ultrasound imaging has been rapidly developed and well-established since the 1950s. The non-invasion, portability, and low costs make it one of the most commonly used imaging techniques in clinical and research settings. In soft bio tissue imaging, for like liver tissue examination and prenatal diagnosis, ultrasound combines deep penetrability and good spatial resolution. The high temporal resolution and non-ionizing-radiation also make the real-time ultrasound efficient in guiding interventional therapy.

In the last two decades, the significant development of ultrasound imaging mainly benefits from the advances in mechanics, electronics, and material science. The piezoelectric micromachined ultrasound transducers [105] and capacitive micromachined ultrasonic transducers have allowed the manufacturing of miniaturized and high-density transducer arrays, which significantly enriched the observable information of sound waves and extended the application scenarios of ultrasound imaging. The ultrasound tomography has dramatically benefited from the dense transducer array, integrated reflection, sound speed, and attenuation modalities in one imaging system for depicting breast parenchyma and fibro-adenoma in detail [106]. The ultrafast ultrasound, powered by plane-wave beamforming, has also ushered in a new ultrasound imaging era. Supported by ultrafast ultrasound, ultrasound localization microscopy collected hundreds of thousands of B-mode (brightness) images enhanced by microbubbles agent in a few minutes and achieved super-resolution ultrasonic angiography (micron-level spatial resolution and centimetre-level penetrability) for Vivo rat brain [107], as shown in figure 13. Continuous volumetric imaging (real-time 3D echography), energized by ultrafast ultrasound and high-density transducer array, can collect 40–50 volumetric images within a single heartbeat cycle been considered as a promising tool for visualization and quantitative assessment of cardiac chamber and valves [108].

The hybrid imaging modalities also attracted much attention in the last decade. Magnetoacoustics imaging, utilizing high-frequency electromagnet wave to excite ultrasonic wave in soft tissues, can get access to high-resolution electrical impedance maps of biomedical tissue [109]. Photoacoustic imaging [110] integrates the advantages of deep penetration from ultrasonic imaging and good resolution from optical imaging and attracted much attention in biomedical imaging society. Researchers also actively fuse ultrasound imaging with other imaging modalities, e.g. electrical impedance tomography [111], MRI [112], and OCT [113], to enrich accessible information for clinical diagnosing and planning.

### Current and future challenges

When the sound waves travel in a tissue, the local sound speed at compressional phases is higher than that at rarefactional phases. Although tissue-harmonic imaging has proved that conventional ultrasound imaging can benefit from this kind of nonlinearity propagation phenomenon, exploring nonlinear acoustics in advanced ultrasound imaging modalities, e.g. contrast-enhanced ultrasonography and real-time 3D echography, is still an open question.

Unlike traditional B-mode imaging, ultrasound computational tomography aims to take into account the full effects of refraction, diffraction, and attenuation of sound waves. However, these effects are usually complicated and hard to decoupled, making the image reconstruction problems of ultrasound computational tomography nonlinear and ill-posed. The reconstruction algorithms, e.g. linear back projection and simultaneous algebraic reconstruction technique, following the basic straight-ray model, are computationally efficient but resulted in crude results. Following some approximation of the Helmholtz wave equation, the advanced reconstruction algorithms produced better results but are computationally intensive. Balancing the speed and accuracy of ultrasound computational tomography is quite challenging.

The plane-wave beamforming is one of the crucial techniques for ultrafast ultrasound imaging. The plane-wave ultrasound transmits in a broad region of interest. It can significantly reduce the number of wave projections for image reconstruction, resulting in ultrafast imaging speed and challenges to the beamforming and reconstruction algorithms. Although some recently published plane-wave beamforming algorithms, e.g. the minimum variance beamforming [114] and short-lag spatial coherence beamforming [115], achieved considerable accuracy, the high computational complexity of these algorithms hampered their real-time applications. Newly developed DL-based beamforming [116] is just beginning. More researches are needed to develop advanced beamforming algorithms.

Furthermore, during the clinical applications of advanced ultrasound imaging techniques, the interpretation of ultrasound images is one of the barrier factors that cannot be ignored. The credible interpretations rely on the clinical specialist with a certain level of expertise. DL-based methods may achieve faster, easier, and high-quality image interpretation. However, the interpretability and generalization of the learning-based methods is still an open question.

### Advances in science and technology to meet challenges

Following Moore's law, computational power has been significantly improved in the last two decades. The flourishing development of HPC techniques, coupled with parallel multi-core compute clusters, cloud computing, edge computing, and large-memory GPU, is removing the barriers between advanced ultrasound imaging and real-time applications.

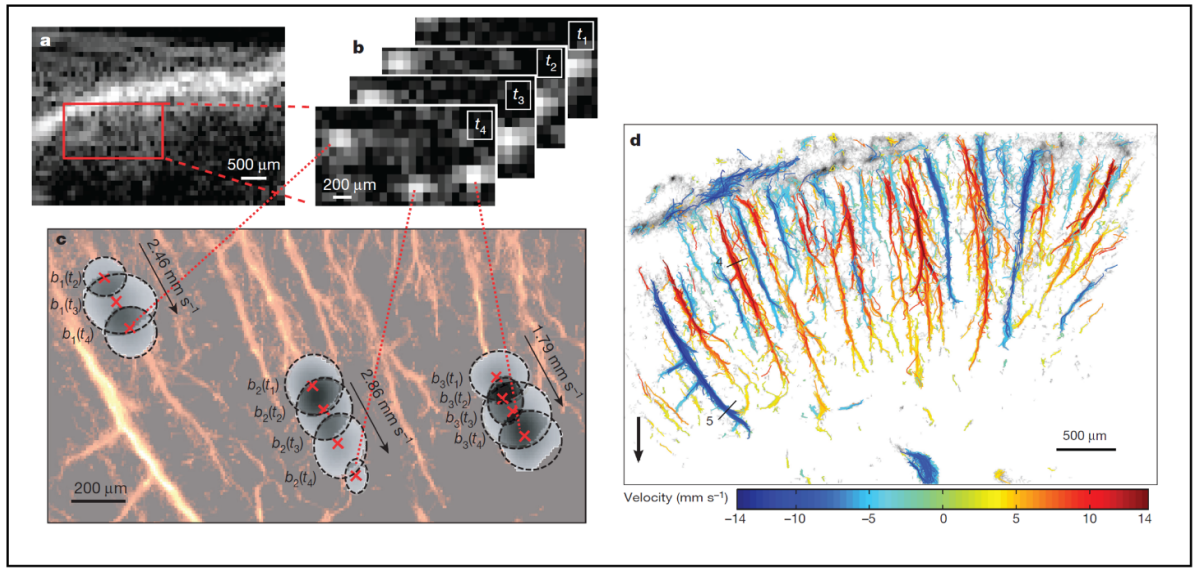

**Figure 13.** Ultrafast detection of individual sources from (a) low-quality B-mode image (averaged stack of 250 beamformed images), through a thinned skull. (b) Four representative frames were separated by 44 ms (t1–t4) and filtered to remove the slow-moving tissue signal. (c) Three independent microbubbles blinking over several milliseconds from (b) were followed in the region of interest within the cortex. The echo of each bubble event (high-contrast pixels) was deconvolved with the PSF to obtain the exact position of the centroid (red crosses). Superposition of thousands of occurrences yields a highly resolved localization map for this region. (d) Dynamic tracking of bubbles separates vessels in two populations with opposite blood flow direction. Reprinted by permission from Springer Nature Customer Service Centre GmbH: Nature. [107] © 2015.

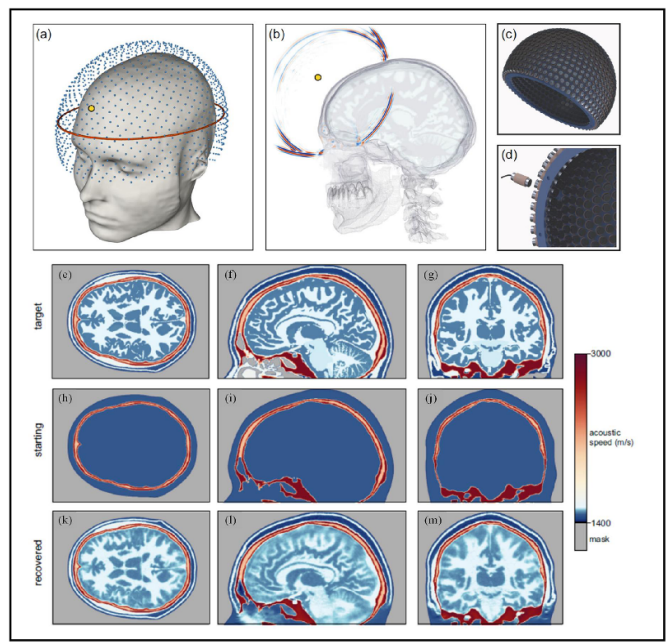

**Figure 14.** (a) Three-dimensional array of transducers used for data generation and subsequent inversion. Each transducer acts as both a source and a receiver. (b) A snapshot in time of the wavefield generated by a source transducer located at the position indicated by the small yellow circle, computed via numerical solution of the 3D acoustic wave equation. The wavefield is dominated by strong reflections from the skull, and by intracranial transmitted energy travelling across the brain. (c) Prototype helmet containing 1024 transducers held rigidly in a 3D-printed framework. (d) Close up of sensor connections in the prototype. (e), (h), (k) Transverse, (f), (i), (l) sagittal, and (g), (j), (m) coronal sections of the images. (e)–(g) Sections from true models. (h)–(j) Sections from starting models. (k)–(m) Sections from recovered model. Reproduced from [117]. CC BY 4.0.

Full-wave inversion, which has been successfully applied in geophysics, recently has shown its potentials in biomedical imaging [117]. Unlike the traditional ray model or approximated wave model, the full-wave inversion method takes account of complex wave propagation in heterogeneous media and multiple scattering of the wave-field. As a result, the total interactions between sound waves and tissues, including refraction, diffraction, and attenuation, are automatically considered in its solution, resulting in a significant spatial resolution improvement. Although the state-of-the-art full-wave inversion requires several minutes for calculating one high-resolution 3D image, as shown in figure 14, its computational cost is hopefully to be significantly reduced by using HPC techniques.

As a new paradigm of HPC, the DL technique provided a convenient solution for learning nonlinear relationships between two data domains. The solution is so efficient that DL technology has revolutionized several research fields, including computer vision, natural language processing, and computational biology. In the last few years, the DL technique also presented promising end-to-end solutions for ultrasound imaging, which could be more robust and flexible than the classical signal or image processing methods. The impacts of DL on the ultrasound imaging involves several aspects, including clutter suppression, beamforming and reconstruction, Doppler signal processing, super-sensitive microbubble localization, acceleration of image reconstruction, and intelligent interpretation of ultrasound images. After being well trained, the DL methods provide efficient models and algorithms beyond conventional imaging and processing techniques, making fast and accurate ultrasound imaging possible.

## Concluding remarks

Ultrasound has played an essential role in biomedical imaging in both clinical and research settings. Advents of high-density transducer arrays and ultrafast ultrasound techniques have significantly enriched the quantity and quality of the information accessible for ultrasound imaging. That also brought new challenges to the research of single and image processing methods in ultrasound imaging society. The advent of powerful parallel computing, full-wave inversion, and DL techniques are broking the bottlenecks of advanced ultrasound imaging techniques. The rapid progress of ultrasound imaging is coming in the near future.

## Acknowledgments

The author acknowledges the support by the National Natural Science Foundation of China (Nos. 61227006, 61571321, 61671322, 81827806, and 61971304).

## 4.6. Signal and image processing methods for biometrics: the impact of DL

*Andreas Uhl*

University of Salzburg

### Status

Biometric recognition is used as alternative to more traditional authentication procedures like tokens or passwords. Unlike the latter two, which base authentication on 'something you have' or on 'something you know', biometrics conduct authentication on 'something you are'. Contrasting to tokens or passwords, biometric identifiers cannot be lost, stolen, forgotten, or passed-on to non-legitimate users. Also, biometric authentication is much more difficult to deceive or misuse, as biometric identifiers are often nearly impossible to duplicate or imitate. Even though there have been attempts to forge biometric traits or to circumvent and/or deceive biometric sensors, also various ways to counteract such fraudulent attacks have been proposed and successfully deployed. Biometric traits can be *physiological*, which depend on the physical properties of a person (e.g. face, fingerprint, iris, various vascular biometrics, ear, palmprint, EEG, and ECG), or *behavioural*, which can be measured during a persons' interaction with the environment (e.g. dynamic signatures, keystroke, gait). Speech is often considered as behavioural, but also carries physiological properties like voice pitch.

The biometric authentication process is a typical pattern recognition procedure and consists classically of several distinct stages: first the data are acquired by the biometric sensor (satisfying certain quality constraints) resulting in the *biometric sample*, and subsequently, this sample is pre-processed to optimally enable the subsequent processing stages (pre-processing may include denoising, contrast enhancement, segmentation, and alignment). The next stage comprises the extraction of biometric features from the sample, resulting in the *biometric template*. The next stage compares the biometric template acquired during authentication to the template stored in the database which was generated during the enrolment process. The determined degree of similarity between the templates is used in the final decision process to decide if sample *verification or identification* has been successful.

Classical biometric recognition is based on hand-crafted or model-based features, derived from the sample data using experts' domain knowledge gained over decades. Contrasting to classical computer vision with its general purpose feature descriptors like local binary patterns or key points along with their descriptors (e.g. SIFT, SURF, HOG, etc), biometric features are often based on intrinsic, trait-specific discriminating properties like face geometry, fingerprint minutiae, or gait patterns. However, with the revolution introducing data-driven and learning-based signal processing, also biometric technology has seen the application of DL [118–123], still with some years delay.

### Current and future challenges

While biometric authentication is a generic pattern recognition process, biometric sample data is by far not generic. Many biometric traits' sample types have quite distinct properties, often significantly different as compared to classical data as processed and used in traditional computer vision applications. This poses a particular challenge for learning-based biometric systems: where do we get our training from? Especially, where to get these data while still respecting privacy-protecting regulations like the European Union GDPR ? In fact, this challenge has impacted significantly the way we perceive biometric technology today: while face recognition was among the mediocre biometric traits in terms of accuracy ten years ago, nowadays it is among the top performers, if not the best. The reason is found in the widespread availability of facial data on-line, in social media platforms of various kinds. It is not a coincidence that companies running social media platforms having access to their users' facial data are able to come up with the most accurate face recognition systems on the market—simply put, they have access to the largest volume of training data to tailor and tune their systems. The advance caused by using DL technology seen in other biometric traits is by far less convincing: often, traditional feature extraction-based technology is able to compete or is still even superior. The lack of sufficient training data and the distinct properties of biometric samples prevent the employment of deep architectures (as required for high accuracies in large populations) and the effective application of TL.

Nevertheless, DL based approaches have been proposed to a large extent for biometric systems. Seen from a distance, subject verification or identification is basically a classification problem. Thus, training a DL based classifier assigning each subject to a dedicated class seems to be a natural choice for biometric recognition. However, there are severe limitations—typically, in biometric systems, new users get enrolled (or eventually deleted), and when doing this, the system has to cope with a dynamically changing number of classes, which limits its applicability. Further, a system with a large number of classes (i.e. subjects) requires a very deep architecture and a significant number of training samples per class—which is often not available. The solution is to train networks to recognise sample similarity/dissimilarity in a Siamese configuration, so that there is no issue in case of unseen subjects.

To overcome the limitations caused by the particular nature of certain biometric sample types (which limits the effectiveness of using networks pre-trained on classical vision datasets) is to employ a combination of TL and fine tuning. Available biometric training data are used to adjust a networks' parameters trained on different types of visual data.

Overall, there is not a single strategy how to apply DL to biometrics. So far, we have discussed the case of using DL as an end-to-end system, thus replacing all processing stages. Contrasting to this, single stages only might be covered by DL technology, while the entire biometric processing chain remains unaltered. For example, this applies to iris recognition in case we apply DL-based semantic segmentation to extract iris texture from samples [124], apply semantic segmentation to extract binary vascular features from finger vein samples [125], or use specifically trained networks to detect fingerprint minutiae [126] or others to conduct latent fingerprint segmentation [127], respectively.

## Advances in science and technology to meet challenges

The unavailability of sufficiently extensive biometric sample training data calls for the development of techniques to address this severe limitation. For example, when combining sample data from different sensor sources, it is by far easier to result in sufficiently sized training data. However, often the sensors are quite different in nature resulting in distinct sample characteristics. This calls for the development and application of multi-sensor and cross-sensor training techniques, involving domain adaptation techniques to compensate for differences also in feature space. Another approach to handle the difficulty as imposed by privacy-protective measures is to synthesize biometric sample data. We have seen techniques being developed that actually generate synthetic samples 'from scratch' obeying natural sample properties, or alternatively, to disentangle identity- and privacy-related information in existing biometric samples such these can be used in GDPR compliant manner.

Another issue with respect to training data has been identified and dealt with recently—it has been discovered that unbalanced training data leads to bias in the recognition process, in particular revealing ethnical and gender-related bias. There-fore, it has turned out that balanced design and fair provisioning of training data are key elements to provide bias-free biometric recognition systems—a fact that again obviously hampers the straightforward establishment of large-scale biometric training data.

Finally, mobile devices get increasingly used as distributed biometric sensor infrastructure. Consequently, also processing of biometric data on these devices becomes imperative, which is particularly challenging in case of DL-based approaches. Thus, advances in this direction are particularly important for future mobile biometric systems.

## Concluding remarks

While DL techniques prevail in many vision areas, their success in biometric recognition system is still impeded by a couple of factors as discussed. However, in case these challenges can be addressed properly, there is hardly doubt that also in this field, DL-based techniques will improve on the state of the art.

The possible usage of DL in biometrics is however not limited to recognition as such. Other popular and promising application areas include presentation attack detection, biometric sensor authentication, learning-based template protection, and many more.

## Acknowledgments

This work has been partially supported by the Austrian Science Fund, Project No. P32201.

# 5. Remote sensing, environmental, and industrial applications

## 5.1. Signal processing for global navigation satellite systems (GNSS)

*Jacek Paziewski*[1] *and Jianghui Geng*[2]

[1] University of Warmia and Mazury in Olsztyn
[2] Wuhan University

### Status

With the progress and modernisation of GNSS including GPS, Galileo, GLONASS, BDS and regional satellite navigation systems such as QZSS and NavIC, now multi-constellation and multi-frequency observations are stimulating innovations [128]. As the most important recent advances, we flag the change from frequency to code division multiple access in GLONASS and the introduction of the third frequency band (L5) in GPS in the latest generation of the satellites. The recent years also witness a great advancement in populating the orbital planes of Galileo and BDS constellations. Therefore, nowadays both systems have become operational to provide positioning, navigation and timing services on a global scale.

This progress induced the scientific community to pay a special attention to the integration of multi-constellation signals into a routine GNSS processing. Also novel positioning models that addressed such integration subsequently emerged. One noticeable recent advance is related to the multi-constellation and multi-frequency precise point positioning with integer ambiguity resolution, which now achieves even a millimetre-level precision at a single receiver [129]. This development would not be possible without significant progress in the precision of GNSS products such as satellite orbits and clocks, and the modelling of the satellite biases that allow isolating carrier-phase ambiguities as integer values [130, 131].

The limitations of low-cost receivers and smartphones have spurred a tremendous effort made by the scientific community to address them. Based on the availability of smartphone-derived GNSS measurements in May 2016, a number of studies on the signal quality and algorithm development aiming at enhancing the positioning precision of mass-market devices have been carried out, which has become recently one of the most frequently investigated topics in GNSS [132]. A common recognition of a poor positioning performance of smartphones and low-cost receivers may soon not hold true, since such receivers are on the way of reaching the performance close to that of survey-grade GNSS receivers.

With these advancements, GNSS is now considered as a fully mature measurement technique which is successfully employed in a number of geoscience applications. Thanks to high altitude of GNSS orbits, satellite navigation signals are comprehensively employed in the water vapour and total electron content sounding, thus contributing to the space weather, meteorology and climate monitoring also coupled with tomographic methods [133]. GNSS reflectometry is a reliable source of information of the geophysical properties of land, water and ice surfaces contributing to the climate research and geohazard monitoring [134].

### Current and future challenges

Notwithstanding a tremendous progress of low-cost GNSS devices and smartphones, we still recognize a number of limitations that deter their application in the most demanding areas of science and technology. The smartphone GNSS antennas suffer from a low and inhomogeneous pattern of gain, high susceptibility to multipath, lack of phase centre models, and a linear polarization that does not prevent the acquisition of the non-line-of-sight left-hand circularly polarized signals [132]. Moreover, users have to handle highly noisy smartphone observations, the presence of unaligned chipset initial phase biases, and other biases that destroy the integer and time-constant properties of carrier-phase ambiguities.

Moreover, a variety of unknown observational biases greatly challenge GNSS data processing since they depend not only on the satellite and receiver hardware, but also on the signal frequencies and code/phase observables. The number of such biases can easily amount to the order of hundreds in a multi-GNSS multi-frequency network analysis, but can hardly be estimated along with other parameters of interest (e.g. positions, clocks, atmosphere, etc) due to rank deficiency. In such cases, these biases will be absorbed by the clock, ionosphere and ambiguity parameters, with the consequence of complicating the time datum definition, the total electron content retrieval, and integer ambiguity resolution. The difficult part in resolving these biases is to separate them from the parameters of interest without harming, but rather enhancing high-precision GNSS.

In addition, the target of an establishment of centimetre-level positioning accuracy within seconds over wide areas has been constantly haunting the GNSS community over the past decades. One of the major obstacles is that GNSS carrier-phase ambiguities cannot be identified as integers within a short period of observations. Multi-GNSS multi-frequency data bring a promising opportunity to form longer wavelength carrier-phase observables of which the ambiguities can be resolved nearly instantly. However, frequency-specific observation errors are adversely amplified therein by nearly a hundred times to a few decimetres, which can hardly allow for the centimetre-level positioning. One such typical error is the multipath effects which are the most difficult to tackle since they are purely environment dependent and prevent the use of any feasible models. Therefore, one of the most critical challenges to speeding up centimetre-level positioning is to minimise the multipath effects within multi-frequency GNSS data.

### Advances in science and technology to meet challenges

While the GNSS satellite technology is developing disruptively, emerging receiver/antenna techniques can also be one of the important routes to resolving the challenges above. The baseband processing to track pseudorange and carrier-phase within GNSS chipsets, which governs how the receiver biases are induced, should be reconciled among a variety of manufacturers. One common recognition among the GNSS community is that such receiver biases should be identical with respect to the satellites emitting the same signals. However, this is not true for GPS/Galileo/BDS satellites, since the receivers' pseudorange biases originate in the chip-shape distortions which are subject to the front-end bandwidths and the correlator designs of receivers. Hence, advanced GNSS baseband processing technologies agreed on by receiver manufacturers are required to minimize such distortions on the pseudo-random code tracking [135].

On the other hand, the multipath effects, especially those jeopardizing GNSS pseudorange, can be potentially reduced by advanced receiver/antenna architecture designs. One can introduce the inertial information of mobile receivers into their baseband chipsets, i.e. the deeply coupled GNSS receivers and inertial measurement units. In this case, a long coherent integration time can be applied to the baseband with the goal of separating the direct and multipath signals in the frequency domain. This idea is predicated on the fact that the direct and multipath signals usually have different Doppler frequencies, which is associated with the different line-of-sight vectors from the satellites to the receiver and from the reflectors to the receiver [136]. Furthermore, an advanced antenna designed with an excellent cutoff pattern can augment the receivers' baseband to better reduce multipath effects. Such a cutoff pattern means that the antenna gain should stay steady over high satellite elevations, but drop sharply when close to local horizons. Though classic choke-ring antennas with clumpy ground planes achieve the steady gain pattern for high elevation satellites, their gain at low elevations is not as good as (about 5–10 dB worse) those ensured by the emerging light and portable helix antennas [137]. Such helix antenna designs and the like could be one of the most achievable directions for the GNSS community to head for in the near future.

## Concluding remarks

In this contribution we have reviewed the current and forthcoming issues in GNSS signal processing. We believe that there are no fundamental barriers that would inhibit a steady improvement in the performance of GNSS positioning and applications. However, the pace of such progress depends on the efficient handling of biases and other error sources in multi-constellation and multi-frequency GNSS signal processing, the quality of satellite products, the magnitude of the research efforts in the processing algorithm development, emerging receiver and antenna technical progress, and fusion with other sensing modalities.

## Acknowledgments

This contribution was supported by the National Science Centre, Poland (No. 2016/23/D/ST10/01546) and National Science Foundation of China (No. 42025401).

## 5.2. Proximal hyperspectral imaging for *in-situ* plants monitoring

*Melanie Ooi*[1], *Ye Chow Kuang*[1] *and Serge Demidenko*[2,3]

[1] University of Waikato
[2] Sunway University
[3] Massey University

### Status

To meet the food needs of the fast-growing world's population, the ability to detect early onsets of plant diseases, predict an optimum harvest time, estimate yield, ensure biosafety and sustainability has become of paramount importance in modern farming.

Plant phenotyping is a term given to the process of measuring a set of physical, biological, and biochemical traits in a plant. With sensors and measurement devices being a ubiquitous part of modern life, their application to *in-situ* plant monitoring and phenotyping combined with the incorporation of information processing is gaining momentum. The associated multidisciplinary field involving engineering, physics, computer science, biology and chemistry, amongst others, is commonly defined as phenomics. Among the application areas of phenomics is the measurement and estimation of the precise nutrient and/or pesticide requirements enabling the evidence-driven trade-off analysis between crop yield and negative environmental impacts. Another important example is the early detection of crop diseases. It facilitates loss-preventive measures thus improving the production yield.

Digital phenotyping activities involve various forms of imaging. One of the most promising of them is hyperspectral imaging, which has been brought from remote to proximal sensing applications allowing simultaneous measurement of structural and spectral responses. Early studies in the field (e.g. [138]) analysed primary characteristics of plants (such as growth, resistance, architecture, physiology, yield, tolerance, etc). Whole plant specimens and their specific components (e.g. roots, leaves, shoots, fruits, seeds, bark, etc) were examined. Recent studies showed that hyperspectral imaging data could be linked to plant chemical composition [139, 140]. It allowed correlating the non-destructive (NDT) imaging assessment with the traditional destructive chemical analytical methods, providing precise measurement of specific chemical compounds. The studies particularly focused on phytochemical compounds related to the crop, i.e. anthocyanin [139], carotenoid, and chlorophyll [140].

The progress in the field has enabled linking the proximal hyperspectral imaging with not only plant phenotypes but also with their genotypes [141]. In turn, this opens up the perspective of identifying genetic strains of the crop with desirable characteristics (e.g. drought tolerance) so to support reliable food production [142].

### Current and future challenges

Traditional challenges of hyperspectral imaging have been associated with acquiring sufficient image data of good quality (i.e. factors such as consistency in imaging setups, illumination, image normalisation, optical distortions, nonlinear sensing sensitivity, etc, contributing to high variations of the plants' spectral response measurements). Besides, some applications demand short acquisition time, high speed of data processing, and image analysis. Current techniques are often not fast or accurate enough [143] while also associated with a high equipment cost.

Some new problems have started to become a bottleneck slowing the progress of hyperspectral imaging for *in-situ* plants monitoring. First of all, it is a lack of commonly accepted protocols for hyperspectral imaging setup and calibration that could ensure repeatable measurements and standardisation in data set structures as well as enable meaningful comparison across experiments. Secondly, image analysis techniques are yet to be urgently advanced to new heights. For instance, image samples of biological specimens are nonhomogeneous even after discounting illumination and imaging noise. This is due to the different spectral characteristics of the specimen parts. Besides, pixel counts of plant parts that are not of interest could overwhelm the pixel counts of the areas of focus of analysis. Efficient annotation and correlation of the measured responses to the intended phenotypes require effective techniques that should be able to identify and separate spectral and/or spatial signatures from surrounding noise.

Challenges in relevant statistical analysis methods are also quite serious. For example, an apparently flat surface of a hyperspectral image represents the measurement of a specimen at varying depths/thickness depending on the penetration of electromagnetic waves of different wavelengths. The confounding effect between thickness (which is often unknown) and the spectral responses poses a problem to standard ML algorithms. Besides, traditional statistical methods do not scale well with large hyperspectral imaging datasets thus requiring particular ML and pattern recognition techniques. Sample sizes are often too small when compared to the variations in crops. This is especially so for mapping of phenotype and genotype [144], thus leading to poor reproducibility and could negate the costs in acquiring, storing and analysing the data.

### Advances in science and technology to meet challenges

A multitude of developments in science and technology is needed to provide further progress towards addressing the challenges of proximal hyperspectral imaging for *in-situ* plants monitoring. The range of such advances is wide: from the sensing and image acquisition platforms to image processing and data analysis and then to specific applications of the information. Several needed advances are outlined below.

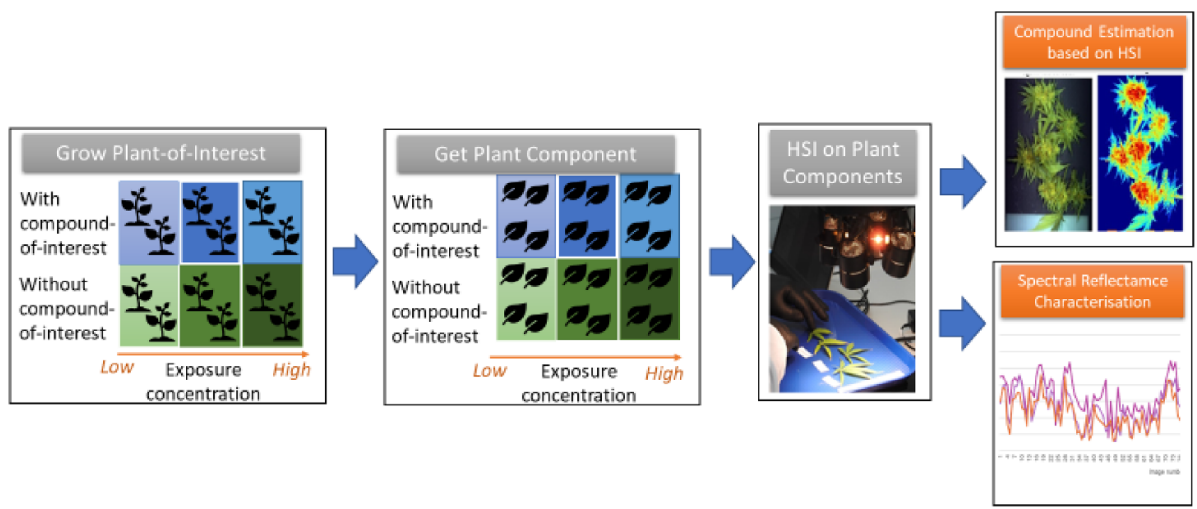

**Figure 15.** Application of proximal hyperspectral imaging to plants.

Advanced toolboxes to fuse multiple sensors aiming to perform complex data analysis are yet to come. Out of all the sensor fusion categories, an efficient algorithm combining a 3D morphological model with hyperspectral images to perform the analysis is acutely lacking. This particular fusion holds great potential to create automated inspection systems with superhuman prediction performance. The creation of the comprehensive toolbox is indeed a very challenging task requiring a team of experts in an array of disciplines such as plant morphology and physiology, machine vision and imaging, sensor fusion, point cloud modelling, ML, and HPC among others.

Statistical ML and pattern recognition have been advancing to address numerous challenges in the niche area of plant phenotyping and hyperspectral images analysis. At the same time, the development of a library of efficient specialised ML and machine vision algorithms to facilitate high-level phenotype analysis would be a very useful common platform of a high value.

Hyperspectral imaging systems are getting more affordable, compact, and resilient. Unfortunately, sometimes the cost improvements are achieved by sacrificing some technical performance characteristics and functions in specific classes of niche applications. In this context, vendors' ability to rapidly customise their produced systems to address specific needs of ever-narrowing niche applications could become a key-important factor to encourage the adoption of proximal hyperspectral technology in agriculture and other relevant fields at an industrial scale.

Robust data management protocols and imaging setups conforming to formulated and adopted good practices would be critical to guarantee the quality of data feeding into the analysis. Reliable data and well-managed databases are prerequisite to enable the use of high-performance data-hungry algorithms associated with image-based plant phenotyping, its applications, and relevant associated areas [145].

## Concluding remarks

Hyperspectral imaging combined with advanced data analytics (figure 15) has been identified and well-accepted as one of the enabling technologies of the Fourth Industrial Revolution addressing the urgent need for accelerating the global food systems transformation. Proximal hyperspectral imaging for *in-situ* plants monitoring is fast coming to the age of maturity, and it has a great potential to become an important part of the quest. At the same time, a significant number of challenging research and development problems associated with algorithms, techniques, and tools for sensing, acquisition, processing, analysis, interpretation, etc, as well as the application of the results are still waited to be addressed [146]. In this section, some of them have been outlined based on the authors' experience gained in the field (e.g. [147]).

## Acknowledgments

This work was supported in part by the Royal Society of New Zealand Te Apārangi through a Rutherford Discovery Fellowship conferred to Melanie Ooi. Furthermore, the authors gratefully acknowledge the valuable discussions with Stefan Hill and Laura Raymond as well as Ray Simpkin, who are senior scientists from two of New Zealand's Crown Research Institutes: Scion and Callaghan Innovation respectively. They are very thankful for the cooperation with staff and the use of facilities that were made available by the partner institutions: (a) Wayne Holmes, Dan Blanchon, Irene Kereama-Royal, and Gregor Steinhorn from Applied Molecular Solutions at Unitec Institute of Technology; (b) Mike Duke and Hin Lim from Robotics, Automation, and Sensing at University of Waikato; (c) Photometric Laboratory at Massey University, (d) Manu Caddie and his team, alongside their horticulture facilities at Rua Bioscience, and for (e) valuable assistance from Sunway University.

### 5.3. Through-wall sensing, signal and image processing

*Francesco Fioranelli*[1] *and Ram M Narayanan*[2]

[1] TU Delft
[2] Pennsylvania State University

#### Status

Through-the-wall (TTW) radar imaging is truly a 'multifaceted technology', requiring a wide breadth of integrated and multidisciplinary knowledge across diverse fields of engineering, such as antennas and array design, signal processing, waveform design, electromagnetic scattering and propagation modelling, and imaging algorithms [148].

Surprisingly, the earliest report of the use of radar signals for through-wall propagation appeared in an advertisement [149]. The first open research papers and surveys on the topic of TTW radar appeared around the late 1980s, attracted by the potential advantages of radar-based through-wall imaging in comparison to alternative technologies based on acoustic and x-ray systems. Radio-frequency electromagnetic waves are attractive since they are able to propagate through common wall construction materials, such as plaster, bricks, concrete, and thick and complex multi-layered structures such as cinder blocks [150].

Although TTW radar systems employing a plurality of waveforms and algorithms have been proposed and validated [148], it is difficult to classify the many different solutions within a unified framework. Performance metrics may include spatial resolution in range and azimuth directions, maximum detection range, stand-off distance, processing time, dynamic range, and practical aspects such as power usage, weight, and size. Alternatively, the amount of information for situational awareness can be used [151], starting from 0D systems indicating human presence and type of human activity, up to 1D, 2D, and 3D systems capable of providing the distance of the TTW targets, their locations in range and angle, and even their comprehensive volumetric signature, respectively. For example, figure 16 shows 3D images of a human target located behind a 60 cm thick reinforced concrete wall wherein the posture of the person, i.e. standing or laying, can be inferred from the 3D images [152].

Despite significant advances, TTW imaging is still an extremely important field for applications in the security and safety domains (e.g. hostage situations, firefighter intervention, building evacuation, earthquake survivor detection, concealed weapon detection) with several open research problems for addressal. Some challenging problems confounding target detection and characterization include non-uniformity of the barrier structure in terms of wall shape and materials, presence of significant indoor clutter, targets of smaller size, and effects of multipath.

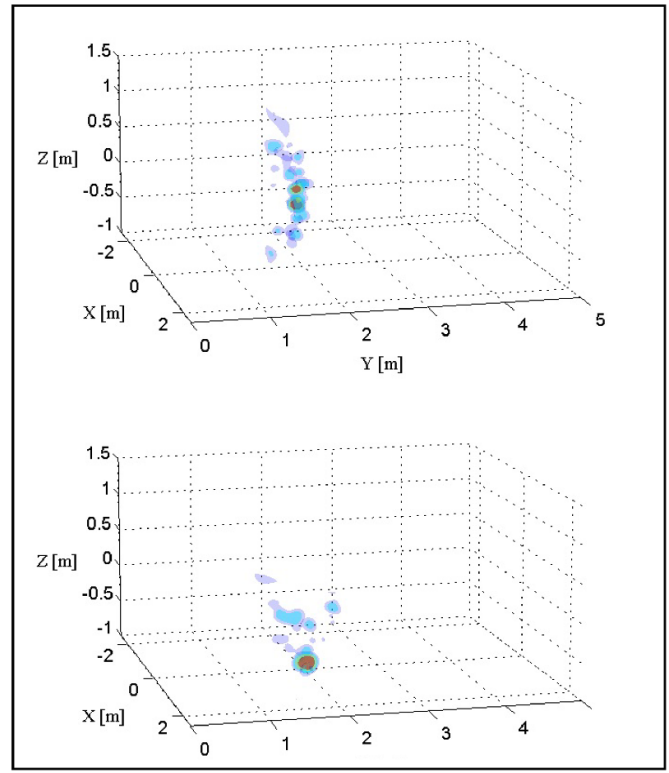

**Figure 16.** TTW measurement scenario and example of results at MS3 Radar Laboratory, TU Delft showing the 3D image of person in standing position (top) and in laying position (bottom) [152]. The front face of the wall is the *XZ*-plane and the radar is positioned towards the left side. Reproduced with permission from [152]. (c) EuMA.

#### Current and future challenges

The wall's effects on the two-way radar propagation path inhibits the formation of clean 2D/3D images. Compensation for wall effects requires knowledge of the dielectric properties of its constituent materials and its internal structure and thickness, which are generally unknown. Approximations based on average values across different construction material types introduce errors in reconstructed image quality. Walls containing internal cavities or metallic rebars create resonant effects extending the wall's signature in time, possibly interfering with those of targets located near the back side of the wall. Another challenge is the mitigation of ghost targets created by multipath reflections from the back and side walls, ceiling, and other obstacles in the indoor environment resulting in artefacts and distortion in the reconstructed images.

The two-way signal attenuation through the wall affects the radar's power link budget and imposes dynamic range constraints for detecting small targets located at far ranges [153]. Furthermore, the strong backscattered reflection at the air–wall interface may saturate the radar receiver. Conventional wall effect removal techniques include subspace projection, spatial filtering, background subtraction, change detection, hardware filtering, and time gating.

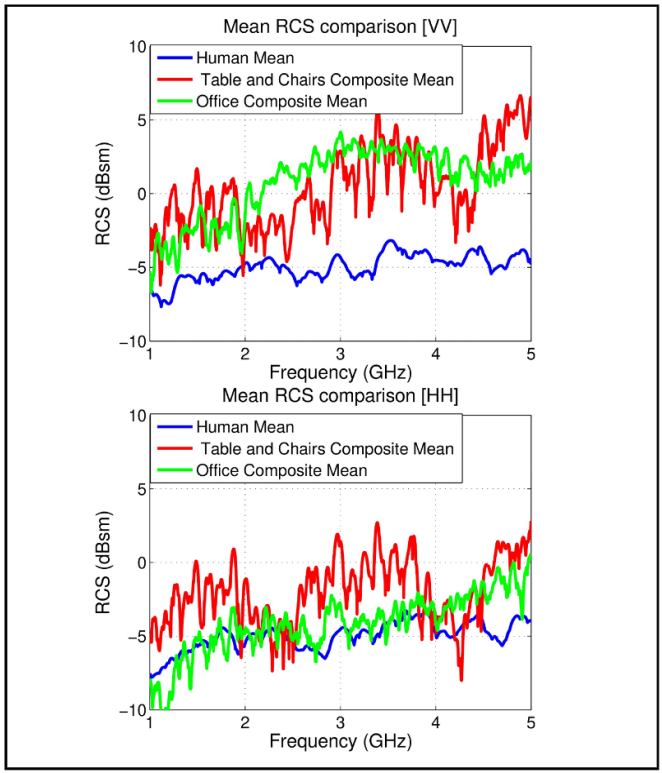

**Figure 17.** Comparison of RCS of human and representative furniture arrangements for vertical–vertical (VV) polarization (top) and horizontal–horizontal (HH) polarization (bottom). © 2015 IEEE. Reprinted, with permission, from [154].

Indoor clutter, typically arising from furniture elements, is another scourge in TTW radar systems causing high backscattered signals compared to those from human targets [154]. High reflections occur from the large planar surfaces of the clutter objects, particularly, the front, sides, and back. Figure 17 shows that the radar cross section (RCS) of furniture arrangements is generally 5 dB higher than that of the human over the 1–5 GHz frequency band. This interferes with and degrades the radar system's ability to accurately detect, locate, and track desired targets. Waveform design exploiting spectral characteristics as features are being investigated for target classification [155].

Achieving high resolution in TTW radar systems is another major challenge. Low frequencies are preferred for better signal penetration through the wall but obtaining the necessary large bandwidth for high down-range resolution is difficult. In addition, antenna beamwidths are wider at lower frequencies inhibiting high cross-range resolution. Techniques such as multiple-input-multiple-output (MIMO) and synthetic aperture radar (SAR) are being investigated for sharpening the cross-range resolution.

## Advances in science and technology to meet challenges

Advances in MIMO radar systems and related signal processing can address some of the open research challenges in

TTW radar imaging [156]. Increasing the number of channels and where possible, given the operational constraints to deploy and synchronize such systems, moving towards forms of distributed, bistatic, or multistatic systems, are expected to enhance situational awareness. Such systems can achieve finer angular resolutions for better characterization of the volumetric target signatures in both azimuth and elevation planes, i.e. finer details of the targets' features and postures for improved situational assessment. Distributed systems can also provide multi-perspective view of the scene of interest, enhancing the perception of certain key target scattering features for identification, and helping discriminate real targets from ghosts [157]. In this context, compressive sensing approaches are additional research paradigms that can play a key role in minimizing the number of physical channels and antennas by exploiting sparse sampling techniques, thereby reducing hardware complexity, expense, size, and processing time.

More sophisticated Doppler processing approaches are needed in the area of human movement and vital sign detection, especially through rubble following disasters [158]. Examples include extraction of very low-level signals due to shadowing and/or propagation through several collapsed walls, rejection of Doppler signals from spurious sources such as fans and motors, and detection of multiple humans of varying sizes.

Intelligent and adaptive waveform design is another open research area where target signatures can be enhanced while suppressing clutter, wall effects, and RF interference. The recent emergence of arbitrary waveform generators and high-speed digital samplers has given rise to several new waveforms and sophisticated pulse-shaping techniques for improved detection and localization of targets behind wall barriers and suppression of ghosts caused by multiple reflections. These include matched illumination, chaotic, orthogonal frequency-division multiplexed, and frequency-modulated interrupted continuous wave (FMICW) waveforms. FMICW waveforms mitigate wall reflections and improve detection of stationary targets and moving or breathing humans behind different wall types [159].

Advances in materials used in antenna fabrication, such as dielectric lenses and metamaterial-inspired superstrates, also provide new directions to significantly enhance TTW radar performance. Examples include bandwidth extension and multi-frequency operation through feeding structure enhancements [160], antenna size reduction using high dielectric constant substrates, and precise beam focusing and control using metasurfaces.

AI and ML techniques can also play a significant role in the performance improvements of TTW systems [161]. Examples include learning better approaches for removal and/or mitigation of wall effects and suppression of ghost targets via generative adversarial networks and denoising networks, as well as dynamic transmit waveform adaptation for improving the signal-to-clutter ratio and switching between different frequency bands to exploit more favourable propagation conditions. As the learning capabilities of these algorithms improve, hybrid techniques could be developed whereby physical prior information on the TTW propagation and scattering can be incorporated into the AI/ML algorithms for target classification and situational awareness, or to provide better estimation of the physical parameters of the wall for deconvolution of its effects.

Traditional TTW radar techniques involve relatively expensive hardware deployment costs and higher power requirement since these operate in an active manner. An emerging area of TTW research involves the exploitation of Wi-Fi signals from the increasing deployment of wireless local area network (IEEE 802.11) technologies in indoor environments due to their ubiquity [162]. Since such systems operate in the passive (i.e. receive-only) mode, opportunities for advances are called for in the development of beamforming and coherent/non-coherent integration schemes to detect and track weak signals from targets, adaptive processing to optimize system performance under uncertainty and noise in the reference channel, and reduction in the computational load for cross-correlation processing.

## Concluding remarks

TTW radar technology is an essential tool in several military, safety and security, law enforcement, commercial, and more recently, in personal applications, and is used primarily to detect humans. Although research in TTW radar has been ongoing for several decades, there are many areas for improvement to enhance target detection, recognition, and imaging. Although the wall signature appears as a nuisance signal, it does provide useful information which may be exploited to suppress its effects in subsequent processing. Detection of targets buried behind irregularly oriented walls formed by building collapses is a major challenge. Most TTW radars are plagued by harsh indoor clutter which may be suppressed by optimization of the transmit spectrum. Future operation TTW radar systems may require a dual-frequency approach wherein lower frequencies are used for high-range detection while higher, perhaps millimetre-wave frequencies, may possess the resolution to detect and resolve targets closer to the wall.

Therefore, next generation measurement systems for through-wall sensing, signal and image processing applications are expected to incorporate and integrate several new technological breakthroughs, such as distributed and dispersed sensors for more complete views of the target scene, intelligent and cognitive waveforms for real-time enhancement of targets and suppression of clutter and interference, advanced Doppler processing techniques for extracting faint vital sign and motion signals, AI and ML approaches for improved feature selection and exploitation, and metamaterial-inspired antennas for reconfigurable beamforming.

## 5.4. Sensors for automated driving

*Carlos Fernandez and Christoph Stiller*

Karlsruhe Institute of Technology (KIT)

### Status

Since early experiments with autonomous vehicles in the 1980s [163] the vision of automated driving has inspired numerous research groups all over the world. Today, modern driver assistance systems are standard to most vehicles and contribute to enhance safety and comfort. Many experts expect the market introduction of fully automated vehicles in the not too far future and that this technology will be disruptive and revolutionary to the automotive industry [164].

Automated vehicles acquire all relevant information through their sensors. For safety reasons, a sensor setup should have redundancy and 360° field of view including different sensor technologies (figure 18).

Automotive radar mainly operates in the 76–81 GHz frequency band. Among all environmental sensors, radar is most robust against atmospheric influences. The strengths of radar include high accuracy (measuring object position and radial velocity) and long range in radial direction, while in azimuthal and vertical accuracy radar remains about an order of magnitude behind cameras and light detection and rangings (lidars).

Automotive lidar operate in the optical spectrum of about 905 nm and measure distance by the time-of-flight principle. Today, lidar sensors provide a set of 3D points in the environment, each associated with a reflectance value. As compared to cameras, lidars are less affected by strong sunlight and robust against spatially variable illumination. Restrictions occur for perception of objects with low reflectivity caused by the material itself or by an unfavourable viewing angle.

Cameras are closest to human perception and operate in the visible spectrum of 380–780 nm. No other sensor principle has comparable perception capabilities when it comes to rich diversity of information. Cameras can detect, classify, measure and track moving and static objects of almost any kind. The status of traffic lights is one example of important information that can hardly be perceived by any other sensor principle.

As automated driving is a safety-critical application, all of the above sensor principles are applied in parallel and appropriate information fusion methods are applied to yield a consistent and plausibilized representation of the environment for subsequent motion planning and control [165].

### Current and future challenges

The abundance of radar, lidar and video sensors as well as the computational power required for sensor data processing result in high cost hindering mass production. Continuous sensor calibration [166] and countermeasures against contamination over the long lifespan of a vehicle are yet other challenges for market introduction. While the redundant use of diverse sensor technologies robustifies perception, some safety-critical perception tasks, such as traffic light state recognition [167] or lane marking recognition, are currently only feasible with a camera, i.e. a single technology.

Another challenge autonomous vehicles may address is occlusion handling. A pedestrian occluded by a bus may cross the road and the vehicle is not able to detect him/her until the pedestrian is inside the detection range. On the other hand, driving under the expectation of pedestrians emerging behind any occluding object would lead to velocities in the range of ∼5 km h$^{-1}$. This dilemma is related to vulnerable road users safety, so it is very important to make improvements in this field for the future.

In the last decade, DL has largely improved perception algorithms. Nevertheless, robustness against different weather conditions must still be improved [168], as in many countries heavy rain and snow are not just exceptional conditions.

Furthermore, DL requires a large amount of representative data to learn from. Driving on the other hand is largely uneventful, i.e. critical situations occur extremely seldom. Therefore, it is likely that a vehicle exposed to a particular critical scenario has only been trained with a moderate number or coarsely related examples and needs to generalize from these. Even common scenarios may differ from one country to another. Infrastructure may also change drastically and traffic lights, traffic signs, road markings and other road elements could have a different appearance that the perception algorithm needs to adapt to.

As an additional 'virtual sensor', high definition maps augment the information basis of autonomous vehicles (AVs) [169]. While these provide information in a large spatial range including traffic rules, road topology, speed limits, etc the main challenge concerning them is how to keep them up to date with the maintenance of the road infrastructure.

### Advances in science and technology to meet challenges

Next generation measurement systems are improving quality and lowering their cost:

- Radar development aims to provide array antennas with improved angular resolution applying beamforming techniques and on-a-chip antennas.
- Likewise, lidar technology migrates from scanning to solid-state technology, removing any moving parts and making sensors cheaper and more robust.
- Cameras are based on silicon technology already and thus follow Moore's law, i.e. their resolution is increased by about a factor 2 every 1.5 years. Emerging from GPUs, low power video data processing units tailored to DL techniques are under development.

Occlusions imply safety critical situations and are very difficult to eliminate from a single vehicle perspective. However, vehicle to vehicle, and vehicle to infrastructure communication may mitigate this problem. In the extreme, when all vehicles intentions or trajectories are shared, more efficient trajectories can be generated and negotiated, where efficiency involves a reduction of traveling time and energy saving.

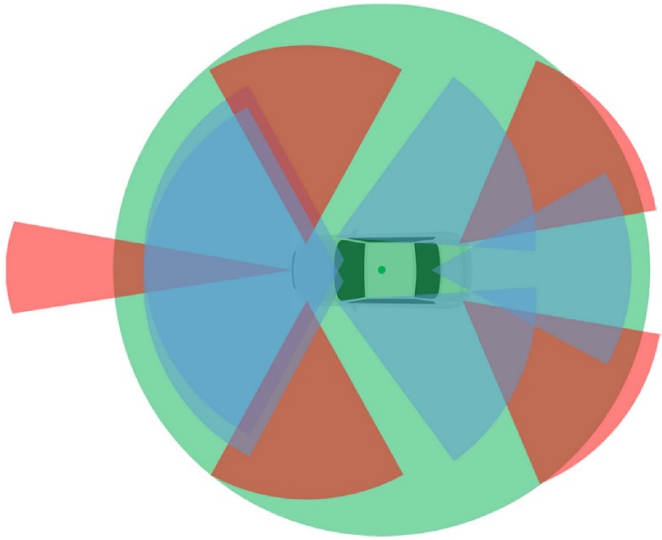

**Figure 18.** Different sensor technologies are usually integrated in typical autonomous vehicle setup. This figure shows an example of how cameras (in blue), RADAR (in red) and LIDAR (in green) are used to increase safety by overlapping sensors' field of view.

Challenging weather conditions such as heavy rain, snow or low light conditions should be considered in vehicle design. Due to the fact that DL requires large amounts of data, dataset generation for a wide variety of purposes becomes a key point. Therefore, tool chains for the acquisition and augmentation of high quality datasets with relevant corner case scenarios gain a growing relevance for many disciplines and applications. Governmental authorities assist such activities as they may lead to homologation requirements and standards.

Datasets covering different regions of the world will contribute to reduce the domain gap and thus improve perception algorithms. Another solution comes from standardization beyond countries to have the same traffic lights, traffic signs, traffic rules, etc and allow algorithms to perform better regardless of the country or city.

Classical signal processing has been replaced or combined with DL techniques. DL is getting integrated in all the software stages. Consequently, a lot of effort is put into understanding deep NNs, what their limitations are and how to provide safety using metrics [170].

To conclude, map verification and crowdmapping is getting popular due to the fact that high level map information requires low bandwidth and it can be transmitted to the cloud.

## Concluding remarks

Automated driving promises a groundbreaking change to our mobility. Thus it comes as no surprise that global players invest great enthusiasm and effort in this technology. These include established vehicle manufacturers that mainly continuously improve their driver assistance systems to adding more and more automation until full automation is reached. In contrast to this evolutionary approach, IT companies aim to build a fully automated vehicle for a limited operational domain in a disruptive approach.

In either approach, sensors and perception algorithms are key components for market introduction. As compared to many other applications, AVs operate in an open world with *a priori* unknown scenarios. Nevertheless, appropriate safety guarantees need to be provided for the perception system before market introduction. Furthermore, safety of the vehicle architecture in general and motion planning in particular are not just mere additional challenges that need to be approached before automated vehicles become reality on our roads.

## 5.5. Acoustic emission (AE) processing for industrial measurement systems

*Konstantina Malamousi*[1], *Spyros Kamnis*[1] *and Konstantinos Delibasis*[2]

[1] University of Thessaly, Dept. of Computer Science and Biomedical Informatics

[2] Castolin Eutectic–Monitor Coatings Ltd, R&D Department

### Status

The scientific era of AEs started in the early 20th century when researchers around the world started to report distinct sounds during the investigation of material deformation. Since then and throughout the years, AE processing has become integral part of industrial sensing technologies aiming to continuously measure critical material properties and to control and report the status of production processes. In principle, AE is a passive monitoring and measurement system with only receivers and without ultrasound transmitters. Currently, the AE uses one or more sensors to 'listen' to a wide range of events with unique NDT and dynamic information characteristics in broadly three application areas: structural testing and surveillance, process monitoring and control, and materials characterisation. The latest state of the art in this field is the development of knowledge-based systems (KBSs) alongside interactive ML techniques. The adoption of advanced industrial measurement technologies, through AE processing, are quickly becoming the first standard for competitive companies and those that fail to do so are likely to be crowded out of an increasingly competitive market. As supply chains become more connected and 'transparent' game changing technologies are shared, through technology transfer licenses, to customers and competitors alike. Such outward transfer of technology has become an important dimension in corporate strategies and has created a highly interdependent ecosystem within most manufacturing sectors. In the absence of sophisticated digital measurement and control technologies, the quality of technical and service support could be extremely compromised. Failure to provide more advanced industrial measurement systems can result in whole batch recalls, downstream production delays, which in turn has a knock-on effect on many sectors such as construction, automotive, aerospace and machinery. The impact extends to costly stock management and even potential loss of contracts. The demand for the development and use of AE technologies is ever increasing. Modern supply chains rely more on digitization as a key enabling technology for building stronger and smarter supply chains through AE measurement and control.

### Current and future challenges

Understanding the physical nature of AE in different processes is the cornerstone in the development of the AE technology. The success and the depth of the technology capabilities depend on the ability to determine unique detectable noise features that can be attributed and effectively correlated to target variables and properties. However, establishing such dependences for different materials and structures is probably the biggest real scientific and technological challenge [171]. To address this critical two main approaches are used. These are broadly the statistical or empirical method and the deterministic or fundamental method. Statistical source characterisation is of wide applicability and it is suitable for complex geometries, anisotropic or inhomogeneous materials. Deterministic characterisation is based on a reasonably detailed knowledge of the physical processes involved in the source so that a mathematical model can be set up. Even though existing techniques are highly efficient, they have also shown some restrictions in handling large number of process parameters and in turn obtaining a judicious relationship between input and output target data. Such difficulties were observed mainly due to the nonlinear and complex nature of phenomena and parameters involved in industrial AE applications. Several other challenges have been identified by [172] in relation to AE. Current AE hardware does not provide sufficient noise immunity against interferences. At the hardware level this requires electromagnetically isolated instrumentation. On the other hand, removing noise during post-processing requires more advanced techniques, as discussed in the next section, alongside outlier detection after frequency filtering. Denoising techniques, such as discrete wavelet filtering of waveform data and Swansong filtering [173] of hit features, are very effective at dividing AE signals into discrete and continuous parts. These can be quantified separately, improving fault diagnosis. Although adaptive filtering has also been applied for segregating specific signals into noise-data components, poor replication and repeatability for generic signals possess a great challenge. In addition, building AE amplifiers, filters, and power supplies with sufficiently low noise levels is much more complicated than for other lower frequency sensing applications (e.g. vibration), due in part to the wide dynamic range and bandwidth required.

### Advances in science and technology to meet challenges

The latest state of the art in noise handling and data analysis is the development of KBSs [174]. KBSs are interactive computer programs which attempt to simulate the existing knowledge and experience-based thought processes and provide a wide range of advice. In this direction, ML techniques with the ability to solve intricate and highly nonlinear processes have attracted worldwide researchers to use them for AE processing optimization, industrial monitoring and control. Different ML techniques such as ANNs and CNNs [175] have been used extensively due to their capability of learning from the past data and using them to predict the target variable more accurately. The basic functionality of the example CNN in figure 19 can be broken down into four key areas. The CNN is trained from AE data collected during thermal spray coating deposition under different spray conditions. As found in other forms of ANN, the input layer comprises of pixel values of a spectrogram image. The convolutional layer determines the output of neurons that are connected to local areas of the input through the calculation of the scalar product of weights and the input volume. The pooling layer then performs down sampling along the spatial dimensionality of the given input. Finally, the fully connected layers perform the same operations found in standard ANNs and attempt to produce class scores. Through this method of transformation, CNNs are able to transform the original input layer by layer using convolutional and down sampling techniques to produce classification scores [176, 177]. The accuracy of the model can be significantly improved when large datasets are used during training. The background industrial noise, form part of the training processes and in this respect special and computationally expensive denoising techniques are not necessary. The demand for development and implementation of automated AE systems with computerized testing involving data acquisition, processing, and evaluation is growing. The true power of ML techniques is harnessed by the synergistic use of measurement data from a probe array composed of multiple sensors.

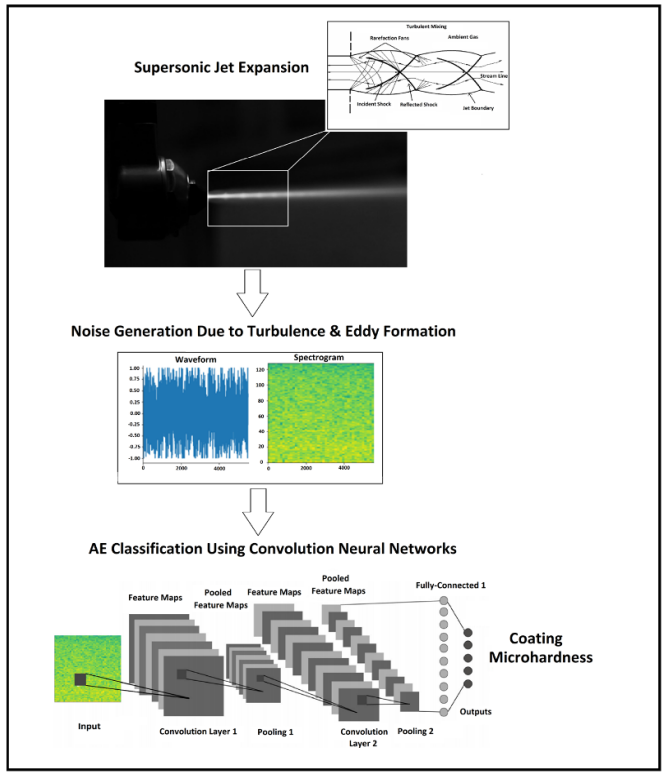

**Figure 19.** Application of AE in surface engineering high velocity thermal spray processes. Measuring the coating microhardness from AE signal classification using convolution neural networks.

## Concluding remarks

AE based systems are flexible technologies that make quality management easier. AE acquired live and historic measurement and process monitoring data enable issues that affect quality to be quickly identified and resolved centrally even across global operations and into the supplier network. New capabilities powered by modern ML techniques, allow for combined real-time data, physical dependency models and intelligence from different platforms. The ability to simulate, measure, predict and improve assets plays a vital role in smart factories of the future [178]. Next generation AE signal processing technologies are expected to drive continuous product improvement and profitability by identifying gaps in performance, diagnosing deficiencies, correcting and reversing negative trends, reducing cost, improving yields, and maintaining equipment reliability. Increased emphasis should be placed on the evolution of ML techniques and hardware including smart, wireless, self-powered sensors that report back AE information through the next generation 5G networks.

## Acknowledgments

The authors would like to acknowledge the support from the UK Research & Innovation (UKRI). Project Grant 132885.

## 5.6. Signal and image processing for condition monitoring and fault diagnosis

*Dong Wang*

The State Key Laboratory of Mechanical Systems and Vibration, Shanghai Jiao Tong University

E-mail: dongwang4-c@sjtu.edu.cn

### Status

Signal and image processing for condition monitoring and fault diagnosis can extract representative features to distinguish different health conditions of systems and critical components and it is always an emerging and cutting-edge topic. Thanks to the rapid development of DL, given sufficiently historical data, advanced fault features can be automatically extracted by using various DL algorithms. Nevertheless, in industrial applications, abnormal and faulty data collected from systems and critical components are seldom available. Most collected data are in a healthy condition. On this condition, most DL algorithms for condition monitoring and fault diagnosis may lose their powerful ability to extract advanced fault features. Further, even though DL algorithms can extract advanced fault features, they are difficult to interpret and cannot be used as objective/loss functions of signal processing and ML algorithms for guiding parameters tuning. As a result, in the field of condition monitoring and fault diagnosis, advances in theoretical investigations on fault feature extraction become extremely important. The first benefit of theoretical investigations on fault feature extraction is that one can know how fault features can be used to quantify the characteristics of signals. The dimensionality of raw signals can be reduced into a few advanced fault features. The second benefit of theoretical investigations on feature extraction is to provide more objective/loss functions for signal processing and ML algorithms. It is well-known that objective/loss functions are one essential component in various algorithms to determine expected properties of output results. Once new objective/loss functions are generated from advanced fault features, an amount of signal processing and ML algorithms will be accordingly improved. The third benefit of theoretical investigations on feature extraction is to serve as advanced health indices to assess degradation performance of systems and critical components. If health indices are with desired monotonicity, trendability, robustness, etc, prognostic models for predicting future feature trends and remaining useful life can be considerably simplified. The fourth benefit of theoretical investigations on feature extraction is to build fault feature databases, which are beneficial to engineers for their practical uses for condition monitoring and fault diagnosis.

### Current and future challenges

The most possible reason why DL algorithms for fault feature extraction is popular is that DL algorithms are easy to implement due to their data-driven black-box behaviours. Once enough historical data are available, any regression/classification models can be fitted to find the relationship between inputs and outputs. However, the interpretation of DL-based features needs strong mathematical skills and wide knowledge. Theoretical investigations on fault feature extraction are parallel to DL-based feature extraction. Some research issues and challenges exist. Firstly, to design suitable features for the quantification of a signal, one must be clear to the distribution characteristics of a signal and needs to design one or more features to characterize the distribution of a signal. Secondly, for condition monitoring and fault diagnosis, the theoretical baseline of features is crucial. This is because the theoretical baseline serves as a reference to know how and when current health conditions deviate from a healthy condition. Thirdly, if multiple features are designed, how to correlate these features and consider their joint condition monitoring and fault diagnosis? Fourthly, varying operating conditions exist during condition monitoring and fault diagnosis. How to design advanced features that are insensitive to varying operating conditions is great of concern. Fifthly, DL algorithms have powerful ability to generate advanced features for condition monitoring and fault diagnosis. How to design advanced features from DL weights is greatly interesting and attractive, which provides the reverse thinking of feature extraction from DL algorithms. Sixthly, a fault feature may only be involved with one designed property to characterize the distribution of a signal. It is known that, in a machine run-to-failure process, signals are varied with degradation time. How to design a composite feature that owns multiple properties for condition monitoring and fault diagnosis is preferable. Seventhly, how to design analytic and simplified expressions for advanced features is beneficial to providing more objective/cost functions for signal processing and ML algorithms. Eighthly, how to design fault feature databases to answer that what kinds of faults need what kinds of fault features is greatly concerned in industrial applications.

### Advances in science and technology to meet challenges

In recent years, advances in theoretical investigations on fault feature extraction to meet fault feature challenges have been reported. Some typical advances are simply introduced as follows. Firstly, for machine condition monitoring, pioneering work is spectral kurtosis [179]. The main idea of spectral kurtosis is to use kurtosis as a fault feature to quantify the impulsiveness of signals preprocessed by a series of band-pass filters to monitor impulsive signals caused by localized machine faults. Moreover, the spectral kurtosis of complex Gaussian noise was calculated as a baseline for machine condition monitoring. Any significant deviation from the baseline can be used to detect machine abnormality. Subsequently, Shannon entropy [180], Gini index [181], smoothness index [182], quasi-arithmetic means [183], correlation dimension and approximate entropy [184], Box–Cox sparsity measures [185], etc have been theoretically investigated and they can be categorised as sparsity measures and complexity measures for machine condition monitoring. In figure 20, signal processing algorithms including WT, squared envelope and Fourier transform, and sparsity measures are linearly decomposed as the weighted sum of physical interpretable nodes and their connections form a NN-like structure [186]. Moreover, the linear decomposition of sparsity measures [187] in figure 21 are incorporated into the NN-like structure to physically interpret the use of hand-crafted features for fault feature extraction and to make the whole NN-like structure physically interpretable. Here, the ReLU function is used because machine condition monitoring is only concerned about positive outputs of the ReLU function to detect machine abnormality. Secondly, composite health indices with desired mathematical properties [188–190] were proposed to fuse process and non-process data to improve the prognostic ability of health indices. Thirdly, besides condition monitoring and fault diagnosis, exponential fault features [191, 192] with random variables and Bayesian updating were proposed to predict remaining useful life of an exponential degradation process. Fourthly, WT was mathematically incorporated into the first layer of CNNs [193] to make CNNs partially interpretable for condition monitoring and fault diagnosis. These works provide preliminary and inspiring results for further theoretical investigations on more advanced fault features for condition monitoring, fault diagnosis and prognostics.

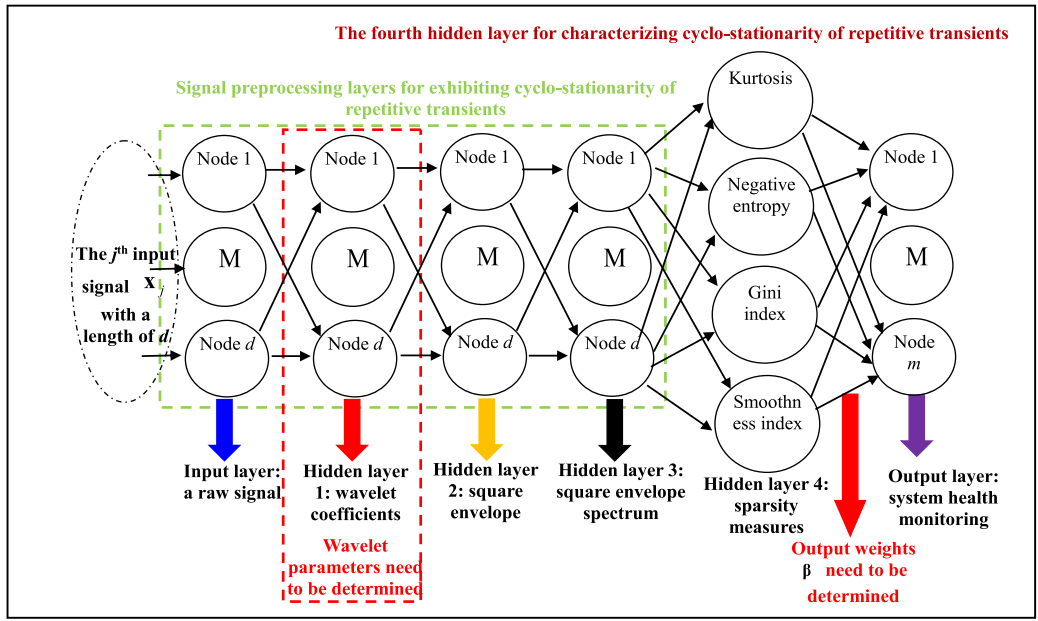

**Figure 20.** Neural network-like structure [186] constructed from the linear decomposition of signal processing algorithms including wavelet transform, squared envelope and Fourier transform, and sparsity measures for machine condition monitoring.

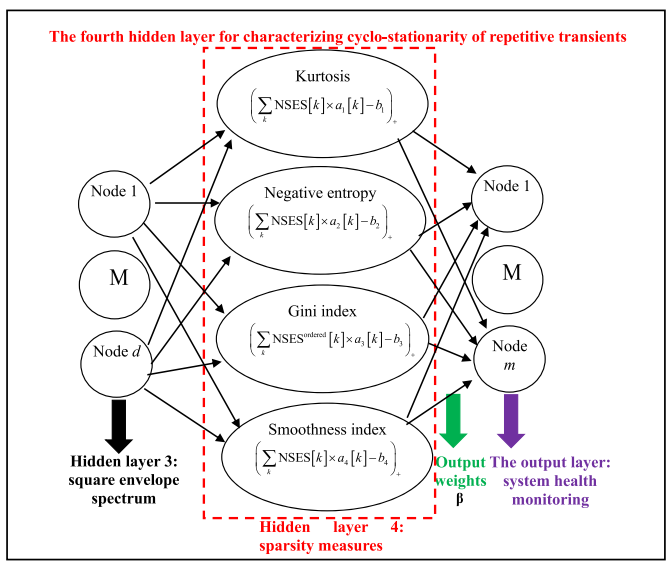

**Figure 21.** Sparsity measures [186] used in the fourth hidden layer of the neural network-like structure for physical interpretation of fault feature extraction.

## Concluding remarks

Hand-crafted features and DL-based fault features are parallel features that can enrich the knowledge domain of condition monitoring, fault diagnosis and prognostics. In recent years, fault features are not fully and theoretically explored and subsequently many fault feature issues and challenges will be addressed. Theoretical investigations on fault features are an emerging research direction that is helpful for engineers and academia to fully understand how popular DL algorithms extract advanced fault features and how to artificially and specifically design advanced hand-crafted features to fully characterize the distribution of a signal. Once various advanced fault features are constructed in fault feature databases, condition monitoring, fault diagnosis and prognostics would become mature and reliable.

**Acknowledgments**

The research work was fully supported by the National Natural Science Foundation of China under Grant No. 51975355. The authors would like to thank five reviewers for their valuable and constructive comments.

## 5.7. Data-driven signal processing for additive manufacturing (AM)

*Jianjing Zhang and Robert X Gao*

Case Western Reserve University

### Status

Over the past decades, AM has increasingly reshaped the manufacturing industry, demonstrating benefits such as higher geometric flexibility and material efficiency as compared to the traditional subtractive processes as represented by metal cutting. While details of the specific AM processes may vary, the general principle of AM is to create tracks of material by fusing powders within a laser-generated melt pool and layering these tracks side-by-side and on top of one another to create 3D components [194]. The sequential relationship of process–structure–property (PSP) is the central paradigm for scientific understanding of AM [195], where the process thermal history involves cycles of repeated melting and solidification of metal powders to produce hierarchical microstructures.

Due to the multi-physics nature of AM processes, significant amount of sensor data can be acquired, making signal processing an indispensable means for data analysis toward the discovery of salient patterns that underlie the PSP relationship. While methods of signal processing have historically been based on analytical models, the increasing availability of data due to ubiquitous sensing has enabled data-driven signal processing as a complementary technique of growing importance [196]. As illustrated in table 1, model-based approaches use analytical representation to perform signal analysis. For example, the characteristic frequency of the temperature signal measured during AM process provides the basis for time–frequency analysis of the thermal history and the inference of properties of the printed part. In comparison, data-driven approaches directly learn salient patterns that are characteristic of the AM process from the measured data, and associates them with part properties by means of numerical modelling.

Besides the availability of data, increasing attention to data-driven signal processing techniques can be attributed to several additional factors: (a) the increasing complexity in AM processes, for which high-fidelity analytical models are not available, (b) new measurement techniques that enable multimodal data collection, and (c) advancement in computational infrastructure for efficient processing of big data [197].

### Current and future challenges

While data-driven signal processing techniques have shown to be effective in deepening the understanding of AM, several challenges have been identified.

*Sensing resolution.* AM process is characterized by rapid interaction between the heat source and metal powders. While sensing systems such as IR camera and pyrometer have enabled the *in-situ* capture of the thermal history associated with the printed part at the macro-scale, knowledge gaps at the meso-scale (e.g. powder scale) remain, such as the mechanism underlying the formation of microstructure (e.g. pore, crack, and keyhole), caused by the limited spatial and temporal resolutions of the existing sensors. Revealing these meso-scale phenomena will provide insight into the influence of the fundamental interactions driving the AM dynamics. Therefore, research of high-resolution sensing system will play a crucial role in advancing the state-of-knowledge in AM [198].

*Image processing.* Due to the spatial and temporal characterizations required for PSP modelling, imaging has become the most prominent sensing modality for AM. However, finding salient patterns embedded in the images that are indicative of process conditions and defects such as porosity, cracking, balling, delamination, and distortion has remained challenging, since each type of the defects exhibits complex geometrical characteristics, surface textures, and varying scales that are difficult to characterize using traditional model-based or even data-driven techniques. Research of advanced image processing techniques is needed to take full advantage of the rich information embedded in the sensing images, which can serve as the technical foundation for establishing the PSP model [199].

*Model interpretation.* The prediction logic of many data-driven methods, especially that of DL models, is generally not transparent to the users. For example, it is unclear which image pixels are used by the network as evidence to recognize process defects and whether findings from the network are consistent with physical knowledge of heat transfer and material formation. These limitations create significant barriers for establishing data-driven signal processing as a trustworthy complement to physics-based reasoning by human experts. Future research on interpretable prediction logic will contribute to establishing user trust in the performance of data-driven and translating the findings into knowledge creation [200].

### Advances in science and technology to meet challenges

In recent years, new advances in science and technology have provided promising means to tackle the challenges in data-driven signal processing in AM.

To provide sufficient spatial and temporal resolutions and capture the fundamental interactions occurred in AM processes, high-speed synchrotron x-ray imaging has been developed to complement the traditional infrared and visible light sensing (figure 22). X-ray imaging enables tracking of the AM dynamics with significantly improved spatial ($\mu$m scale instead of mm scale) and temporal ($10^7$ Hz as compared to $10^5$ Hz) resolutions [198].

Recent advancement in data-driven techniques such as CNNs, which are inspired by the mechanism of human vision system for image processing [199], has shown to be effective in analysing and extracting salient features from the complex image compositions to support PSP modelling. In [201], a CNN-based method for recognizing surface defects induced by process condition deviations has been developed, where a bi-stream CNN structure has been designed to analyse visible light images of both AM part slices and powder layers (figure 23) and fuse the extracted defect patterns from both to enable comprehensive evaluation of the AM condition. Experimental evaluation has shown high condition recognition accuracy of 99.4%.

**Table 1.** Representative methods for model-based and data-driven signal processing.

| Approach | Method | Representation |
|---|---|---|
| Model-based | Wavelet transform | $\mathrm{cwt}(s,\tau) = \frac{1}{\sqrt{s}} \int x(t)\,\psi^* \left(\frac{t-\tau}{s}\right) \mathrm{d}t$ |
| | Kalman filter | $p\left(\boldsymbol{x}_k \vert \boldsymbol{z}_{k-1}\right) \sim N\left(\boldsymbol{x'}_k + K\left(\boldsymbol{z}_k - G_k \boldsymbol{x'}_k\right), P'_k - KG_k P'_k\right)$ |
| Data-driven | Support vector machine | $\mathrm{argmin}_{\boldsymbol{w},\boldsymbol{b}} \frac{\|\boldsymbol{w}\|^2}{2}, \text{ s.t. } y_i \left(\boldsymbol{w}^T \boldsymbol{x}_i + b\right) \geq 1$ |
| | Deep neural networks | $\mathrm{argmin}_{\boldsymbol{w}} - \sum \boldsymbol{y}_i \log\left(\sigma\left(f_{\boldsymbol{w}}\left(\boldsymbol{x}_i\right)\right)\right)$ |

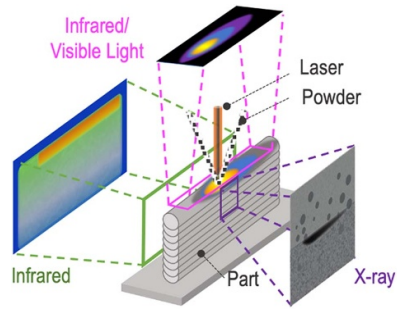

**Figure 22.** Imaging in infrared, visible light and x-ray range for AM process monitoring.

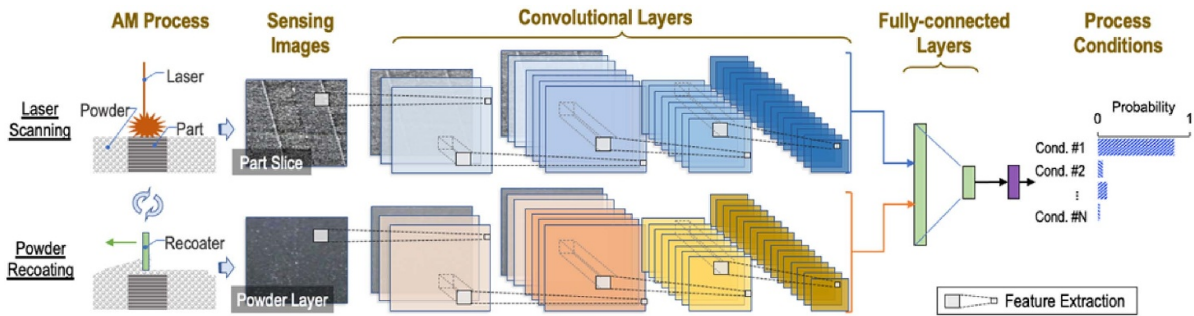

**Figure 23.** Bi-stream CNN structure for process condition monitoring [201]. Reprinted from [201], Copyright (2019), with permission from Elsevier.

Research efforts towards improving model interpretability have also been reported, which fall under two categories. The first quantifies the relevance of individual image pixels in network decision-making as part of a post-analysis. A representative technique is the layer-wise relevance propagation [202], which redistributes network's prediction backwards until a score is assigned to each image pixel. Positive score values indicate that the corresponding pixels are used as evidence by the network to determine process conditions. In comparison, techniques in the second category embeds knowledge of process physics directly into the model design to improve its interpretability and consistency with physics. For example, WaveletKernelNet has been developed [193], in which a continuous wavelet convolutional layer is designed to enable the discovery of physically interpretable kernels for feature extraction.

## Concluding remarks

With the increasing availability of measurement data, data-driven signal processing has developed into a promising alternative to complement traditional signal processing techniques, with the advantage of directly learning salient patterns that underlie the data to understand the physical phenomenon of interest. Using AM as an application scenario, this section summarizes the state of data-driven signal processing for manufacturing process monitoring. In addition, major challenges to improving sensing resolution, image processing and model interpretability are summarized, and early research efforts in tackling these challenges are highlighted.

## Data availability statement

The data that support the findings of this study are available upon reasonable request from the authors.

## ORCID iDs

Dimitris K Iakovidis  https://orcid.org/0000-0002-5027-5323

Melanie Ooi  https://orcid.org/0000-0002-1623-0105

Ye Chow Kuang  https://orcid.org/0000-0002-5423-9653

Serge Demidenko  https://orcid.org/0000-0001-9883-9311

Vladimir Sinitsin  https://orcid.org/0000-0002-6049-9830

Manus Henry  https://orcid.org/0000-0002-9677-1234

Andrea Sciacchitano  https://orcid.org/0000-0003-4627-3787

Stefano Discetti  https://orcid.org/0000-0001-9025-1505

Silvano Donati  https://orcid.org/0000-0002-2977-0194

Michele Norgia  https://orcid.org/0000-0002-8571-1527

Ilias Maglogiannis  https://orcid.org/0000-0003-2860-399X

Selina C Wriessnegger  https://orcid.org/0000-0003-4345-7310

Anthony H Aletras  https://orcid.org/0000-0002-3786-3817

Feng Dong  https://orcid.org/0000-0002-8478-8928

Shangjie Ren  https://orcid.org/0000-0003-2220-3856

Jacek Paziewski  https://orcid.org/0000-0002-6033-2547

Francesco Fioranelli  https://orcid.org/0000-0001-8254-8093

Ram M Narayanan  https://orcid.org/0000-0003-3568-2702

Dong Wang  https://orcid.org/0000-0003-4872-4860